\documentclass[journal]{IEEEtran}
\usepackage{epsfig} % for postscript graphics files
\usepackage{mathptmx} % assumes new font selection scheme installed
\usepackage{times} % assumes new font selection scheme installed
\usepackage{amsmath} % assumes amsmath package installed
\usepackage{amssymb}  % assumes amsmath package installed
\usepackage{epstopdf}
\usepackage{supertabular}
\usepackage{subfigure} %JYM: subfigures
\usepackage{graphicx}
\usepackage{color}

\newtheorem{Theorem}{{\bf Theorem}}

\newtheorem{Proposition}{{\bf Proposition}}

\newtheorem{definition}{{\bf Definition}}
\newtheorem{Lemma}{{\bf Lemma}}

\newtheorem{Corollary}{{\bf Corollary}} %JYM: Collorary
\newcommand{\nop}[1]{} %JYM: This new commend allows to comment text within "{}" in "\nop{}".

\ifCLASSINFOpdf
\else
\fi
\begin{document}
\title{Approximate Consensus Multi-Agent Control Under Stochastic Environment with Application to Load Balancing}

\author{Natalia~Amelina, %~\IEEEmembership{Member,~IEEE,}
        Alexander~Fradkov,~\IEEEmembership{Fellow,~IEEE,}
        Yuming~Jiang,~\IEEEmembership{Member,~IEEE,}
        and~Dimitrios~J.~Vergados,~\IEEEmembership{Member,~IEEE}% <-this % stops a space
\thanks{N. Amelina is with the Department of Telematics, Norwegian University of Science and Technology, NO-7491, Trondheim, Norway and also with Faculty of Mathematics and Mechanics, St. Petersburg State University, 198504, Universitetskii pr. 28, St. Petersburg, Russia, e-mail: natalia\_amelina@mail.ru.}% <-this % stops a space
\thanks{A. Fradkov is with the Institute of Problems in Mechanical Engineering, 199178, St. Petersburg, Russia and also with Faculty of Mathematics and Mechanics, St. Petersburg State University, 198504, Universitetskii pr. 28, St. Petersburg, Russia  e-mail: fradkov@mail.ru.}% <-this % stops a space
\thanks{Y. Jiang is with the Department of Telematics, Norwegian University of Science and Technology, NO-7491, Trondheim, Norway, e-mail: jiang@item.ntnu.no.}% <-this % stops a space
\thanks{D.~J. Vergados is with the Department of Telematics, Norwegian University of Science and Technology, NO-7491, Trondheim, Norway, e-mail: dimitrios.vergados@item.ntnu.no.}% <-this % stops a space
\thanks{Some of the results were presented at IEEE MSC Conference 2012, SNPD Conference 2012, and ICN Conference 2013.}} %\cite{Amelina12, Amelina12SNPD, Amelina13ICN}

\maketitle

\begin{abstract}
%\boldmath
The paper is devoted to the approximate consensus problem for networks of nonlinear agents with switching topology, noisy and delayed measurements. In contrast to the existing stochastic approximation-based control algorithms (protocols), a local voting protocol with nonvanishing step size is proposed. Nonvanishing (e.g., constant) step size protocols give the opportunity to achieve better convergence rate (by choosing proper step sizes) in coping with time-varying loads and agent states. The price to pay is replacement of the mean square convergence with an approximate one. To analyze dynamics of the closed loop system, the so-called method of averaged models is used. It allows to reduce analysis complexity of the closed loop system. In this paper the upper bounds for mean square distance between the initial system and its approximate averaged model are proposed. The proposed upper bounds are used to obtain conditions for approximate consensus achievement.
%JYM

%The paper is devoted to the approximate consensus problem for networks of nonlinear agents with switching topology, noisy and delayed measurements. In contrast to the existing stochastic approximation-based control algorithms (protocols), local voting protocols with nonvanishing step size are proposed.  Nonvanishing (e.g., constant) step size allows to achieve better convergence rate and copes with time-varying loads and agent states. The price to pay is replacement of the mean square convergence with an approximate one. To analyze dynamics of the closed loop system, the so-called the method of the averaged models is used. It allows to reduce complexity of the closed loop system analysis. In this paper new upper bounds for mean square distance between the initial system and its approximate average model are proposed. The proposed upper bounds are used to obtain conditions for approximate consensus achievement.

%The method is applied to the balancing problem of information capabilities in stochastic dynamic sensor network with incomplete information about the current state of nodes and changing set of communication links. This problem is reformulated as consensus problem in noisy model with switched topology.

The method is applied to the load balancing problem in stochastic dynamic networks with incomplete information about the current states of agents and with changing set of communication links. The load balancing problem is formulated as consensus problem in noisy model with switched topology. The conditions to achieve the optimal level of load balancing (in the sense that if no new task arrives, all agents will finish at the same time) are obtained.
%JYM

%The conditions to achieve the optimal level of agents load are obtained.
The performance of the system is evaluated analytically and by simulation. It is shown that the performance of the adaptive multi-agent strategy with the redistribution of tasks among ``connected'' neighbors is significantly better than the performance without redistribution. The obtained results are important for control of production networks, multiprocessor, sensor or multicomputer networks, {\it etc.}
%JYM

%The performance of the system is evaluated analytically and by simulations of simultaneously processing of $10^6$ tasks by $1024$ agents with $2048$ links. It is shown that the performance of the adaptive multi-agent strategy with the redistribution of tasks among ``connected'' neighbors is significantly better than the performance of  the strategy without redistribution. Obtained results are important for control of production networks, multiprocessor, sensor or multicomputer networks, {\it etc.} %JYM

\end{abstract}


% \fi
\IEEEpeerreviewmaketitle

\section{Introduction}
\IEEEPARstart{T}{he} problems of control and distributed interaction in dynamical networks attracted much attention in the last decade. A number of survey papers \cite{OlfatiFaxMurray07, RenBeardAtkins07}, monographs \cite{Wu2007, Ren, BulloCortezMartinez07}, special issues of journals \cite{IEEE1, IEEE2, IEEE3} and edited volumes \cite{Armbruster} have been published in this area. This interest has been driven by applications in various fields, including, for example, multiprocessor networks, transportation networks, production networks,  coordinated motion for unmanned flying vehicles, submarines and mobile robots, distributed control systems for power networks, complex crystal lattices, and nanostructured plants  \cite{OlfatiFaxMurray07, RenBeardAtkins07, Wu2007, Ren, BulloCortezMartinez07, IEEE1, IEEE2, IEEE3, Armbruster, GranichinScobelev12}. %JYM
%The stochastic gradient-like (stochastic approximation) methods have also been used in the presence of stochastic disturbances and noise \cite{Huang, Olfati, Ren2, Tsitsiklis, Huang2, Li2}. %JYM

Despite a large number of publications, satisfactory solutions have been obtained mostly for a restricted class of problems (see  \cite{OlfatiFaxMurray07, RenBeardAtkins07, Wu2007, Ren, BulloCortezMartinez07, IEEE1, IEEE2, IEEE3, Armbruster} and references therein). Factors such as nonlinearity of agent dynamics, switching topology, noisy and delayed measurements may significantly complicate the solutions. Additional important factors can be the limited transmission rate in the channel and discretization phenomenon. In the presence of various disruptive factors, asymptotically exact consensus may be hard to achieve, especially in a time-varying environment. For such cases, approximate consensus problems should be examined.   %JYM

{\em In this paper, we investigate the approximate consensus problem in a multi-agent stochastic system with nonlinear dynamics, measurements with noise and delays, and uncertainties in the topology and in the control protocol}. Such a problem is important for the control of production networks, multiprocessors, sensor or multicomputer networks, {\it etc.}  {\em As an example, the load balancing system in a network with noisy and delayed information about the load and with switched topology is studied.} In contrast to the existing stochastic approximation-based control algorithms (protocols), local voting with {\bf nonvanishing step size} is considered. %JYM
%In addition, numerous articles are devoted to the load balancing problem (e.g., \cite{Friedrich, Li, Kechadi, Cybenko89, Yawei, Gilly, Sim}), indicating the relevance of this topic. However, most of these articles usually do not consider noise and delays. Within a single computer setting, their such consideration could be rather realistic. However, if we consider networked systems, the effect of noise, delays and link-``breaks'' should be taken into account.

In the literature, the average consensus problem on graphs with noisy measurements of its neighbors' states, under general imperfect communications is considered in \cite{Huang2, Rajagopal11}, where stochastic approximation-type algorithms with {\em decreasing to zero step size} are used. Noisy convergence with nonvanishing step-size was studied in~\cite{Guo}, but the control step parameters were chosen differently for different agents and the considered network scenario is a specific one. The stochastic gradient-like (stochastic approximation) methods have also been used in the presence of stochastic disturbances and noise \cite{Olfati, Ren2, Tsitsiklis, Huang2, Li2}. However, in these works \cite{Huang2, Guo, Olfati, Ren2, Tsitsiklis, Li2}, the considered network scenarios are often specific ones, much simpler than the more general scenario considered in this paper. %JYM

In \cite{Huang}, the considered network scenario is most close to that in this paper. In \cite{Huang}, a stochastic approximation type algorithm was proposed for solving consensus problem and justified for the group of cooperating agents that communicate with imperfect information in discrete time, under the conditions of dynamic topology and delay. Under some general assumptions a necessary and sufficient condition was proved for the asymptotic mean square consensus when {\em step size tends to zero} and with a simple form of dynamic functions: $f^i(x_t^i, u_t^i)=u_t^i$ in the paper in~\cite{Huang}. However, as to be shown in the results, under dynamic state changes for the agents (e.g., feeding new jobs), using step sizes that decrease to zero may greatly affect the convergence rate. In our paper, we consider a more general case of functions $f^i(x_t^i, u_t^i)$ and step size $\alpha_t$ nondecreasing to zero.  %JYM

In \cite{Granichin, Vakhitov, Granichin2, Borkar08} the efficiency of stochastic approximation algorithms with constant step size was studied for some specific cases with different properties and constraints than these considered in this paper.  %JYM

As for the load balancing problem, numerous articles are devoted to it (e.g., \cite{Friedrich, Li, Kechadi, Cybenko89, Yawei, Gilly, Sim}), indicating the relevance of this topic. However, most of these articles do not consider noise or delays. While in a single computer this assumption could be rather realistic, if we consider networked systems, noise, delays and possible link-``breaks'' need to be justified. The load balancing problem in centralized networks with noisy information about load and agent productivity was analyzed in \cite{Granichin10, Vakhitov11}. In such a centralized network, there is a load broker that redistributes jobs among agents. However, in case when each agent is not connected with every other agent, it is not possible to choose one of the agents as the load broker, who would distribute the jobs among the agents. In this case, it is necessary to consider decentralized networks. However, to the best of our knowledge, few results for load-balancing in such distributed networks are available. %JYM

In this paper, the results of our previous works~\cite{Amelina12, Amelina12AiT, Amelina12SNPD, Amelina12Vest, Amelina13ICN} are summarized, extended and improved. %In particular, we relax the conditions for the results. In addition, new and much larger size simulation experiments were performed and results added.  %JYM
In particular, we relax the assumption of the weights boundedness of the control protocol, replacing it by the boundedness of its variances. In addition, new and much larger size simulation experiments were performed and results added.
  %JYM

The contributions of the paper are several-fold. First, the approximate consensus problem for a general network scenario is investigated, which is a network of nonlinear agents with switching topology, noisy and delayed measurements. Second, in this approximate consensus problem, we specifically consider a more general state function $f^i(x_t^i, u_t^i)$ and step size $\alpha_t$ nondecreasing to zero in the local voting protocol. Third, to analyze the dynamics of the stochastic discrete systems, the method of averaged models (Derevitskii-Fradkov-Ljung (DFL)-scheme) \cite{DF74ARC, Ljung77, Kushner77} is adopted. Forth, the consensus conditions for the case without delays in measurements and for the case with delays are obtained. In addition, to demonstrate the use of the obtained results, the load balancing problem in a distributed network is studied. Furthermore, simulation results validating the analysis are presented.  %JYM

The rest of the paper is organized as follows. In Section II, the basic concepts of graph theory are introduced, the consensus problem is posed, and some preliminary results for consensus conditions in non-stochastic system are considered. In Section III, the basic assumptions are described and the consensus conditions for the case without delays in measurements and for the case with delays are obtained. In Sections IV, the load balancing problem is considered, and analytical and simulation results are presented and discussed. Section V contains conclusion remarks. %JYM

%%%%%%%%%%%%%%%%%%%%%%%%%%%%%%%%%%%%%%%%%%%%%%%%%%%%%%%%%%%%%%%%%%%%%%%%%%%%%%%%
\section{Preliminaries}

%%%%%%%%%%%%%%%%%%%%%%%%%%%%%%%%%%%%%%%%%%%%%%%%%%%%%%%%%%%%%%%%%%%%%%%%%%%%%%%%
\subsection{Concepts of Graph Theory}

First we present the notation used in this article. The agent index is used as a superscript and not as an exponent.
%Let $[x_1;\ldots;x_l]$ denotes the column vector obtained from numbers $x_1,\ldots,x_l$.

%by vertical concatenation of the $l$ vectors.

Consider a network as a set of agents (nodes) $N = \{ 1, 2, \ldots, n\}$.

A \textit{directed graph (digraph)} $G=(N,E)$ consists of a set $N$ and a set of directed edges $E$. Denote the \textit{neighbour set} of node $i$ as $N^{i} =\{ j: (j,i) \in E\}$.

We associate a weight $a^{i, j}>0$ with each edge $(j, i)\in E$. Matrix $A =[a^{i, j}]$ is called an \textit{adjacency or connectivity matrix}  of the graph. Denote ${\cal G}_A$ as the corresponding graph. The \textit{in-degree} of node $i$ is the number of edges having $i$ as head. The \textit{out-degree} of node $i$ is the number of edges having $i$ as tail. If the in-degree equals to the out-degree for all nodes $ i \in N$ the graph is said to be \textit{balanced}.
Define the \textit{weighted in-degree} of node $i$ as the $i$-th row sum of $A$: $d^i(A)=\sum_{j=1}^n a^{i, j}$ and $D(A)={\rm diag}\{d^1(A), d^2(A), \ldots, d^n(A)\}$ is a corresponding diagonal matrix. The symbol ${\cal L}(A)=D(A)-A$ stands for the \textit{Laplacian} of graph ${\cal G}_A$.

\textit{A directed path} from $i_{1} $ to $i_{s} $ is a sequence of nodes $i_{1}, \ldots, i_{s}, \; s \ge 2$, such that $(i_{k} ,i_{k+1} )\in E, k \in \{1,2, \ldots, s-1\} $. Node $i$ is said to be \textit{connected} to node $j$ if a directed path from $i$ to $j$ exists. \textit{The distance} from $i$ to $j$ is the length of the shortest path from $i$ to $j$. The graph is said to be \textit{strongly connected} if $i$ and $j$ are connected for all distinct nodes $i,j \in N$.

A \textit{directed tree} is a digraph where each node $i$, except the root, has exactly one parent node $j$ so that $(j,i) \in E$. We call $\overline{{\cal G}_A}=(\overline{N},\overline{E})$ a \textit{subgraph} of ${\cal G}_A$ if $\overline{N} \subset N$ and $\overline{E} \subset E \cap \overline{N}\times \overline{N}$. The digraph ${\cal G}_A$ is said to contain a \textit{spanning tree} if there exists a directed tree ${\cal G}_{tr}=(N,E_{tr})$ as a subgraph of ${\cal G}_A$.

The following fact from graph theory will be important.
\begin{Lemma} \cite{Ren2, Agaev00}
The Laplacian ${\cal L}(A)$ of the graph ${\cal G}_A$ has rank $n-1$ if and only if the graph ${\cal G}_A$ has a spanning tree.
\end{Lemma}

%JYM: This corollary is a direct part of the above lemma and hence is redundant.
%We note an important corollary:
%\begin{Collorary} If the graph ${\cal G}_A$ has a spanning tree, then the Laplacian ${\cal L}(A)$ has rank $n-1$.
%\end{Collorary}

The symbol $d_{\max}(A)$ denotes a maximal in-degree of the graph ${\cal G}_A$. In correspondence with the Gershgorin Theorem \cite{Gershgorin}, we can deduce another important property of the Laplacian:
{\it all eigenvalues of the matrix ${\cal L}(A)$
have nonnegative real part and belong to the circle with center on the real axis at the point $(0, d_{\max}(A))$ and with radius which equals to $d_{\max}(A)$.}

Let $\lambda_1, \ldots, \lambda_n$ denote eigenvalues of the matrix ${\cal L}(A)$. We arrange them in ascending order of real parts: $0 \leq Re(\lambda_1) \leq Re(\lambda_2) \leq \ldots \leq Re(\lambda_n)$.
By virtue of Lemma 1, if the graph has a spanning tree then $\lambda_1=0$  is a simple eigenvalue and all other eigenvalues of ${\cal L}$ are in the open right half of the complex plane. %JYM

The second eigenvalue $\lambda_2$ of matrix ${\cal L}$ is important for analysis in many applications. It is usually called Fiedler eigenvalue. For undirected graphs it was shown in \cite{Wu2007} that:
\begin{equation}
Re(\lambda_2) \leq \frac{n}{n-1} \min_{i \in N} d^i(A),
\end{equation}
and for the connected undirected graph $G_{A}$
\begin{equation}
Re(\lambda_2) \geq \frac{1}{{\rm diam} G_{A} \cdot {\rm vol} G_{A}},
\end{equation}
 where ${\rm diam} G_{A}$ is the longest distance between two nodes and  ${\rm vol} G_{A} = \sum_{i \in N}d^i(A)$.

For all vectors the $\ell_2$-norm will be used, i.e. a square root of the sum of all its elements squares.

For reader's convenience, we provide a list of key notation used in this paper.

\begin{center}
\begin{supertabular}{|p{1.3cm}|p{6.4cm}|}
$N$ & $\{ 1, 2, \ldots, n\}$ --- the set of nodes \\
$E$ & $\{i, j\}$ --- the set of edges, $i, j \in N$ \\
$a^{i, j}$ & weight of edge $(j, i)\in E$ \\
$(N, E)$ & digraph with nodes $N$ and edges $E$ \\
$N^{i}$ & neighbour set of node $i$ \\
$A$: &  adjacency or connectivity matrix\\
${\cal G}_A$ & graph defined by the adjacency matrix $A$ \\
$d^i(A)$ & $\sum_{j=1}^n a^{i, j}$ --- weighted in-degree of node $i$ ($i$-th row sum of $A$) \\
$d_{\max}(A)$ & maximal in-degree of the graph ${\cal G}_A$ \\
$D(A)$ & ${\rm diag}\{d^1(A), d^2(A), \ldots, d^n(A)\}$ --- diagonal matrix of waighted in-degree of $A$  \\
${\cal L}(A)$ & $D(A)-A$ --- Laplacian of graph ${\cal G}_A$ \\
$\lambda_1, \ldots, \lambda_n$ & eigenvalues of the matrix ${\cal L}(A)$ \\
${\rm diam} G_{A}$ & diameter, the longest distance between two nodes \\
${\rm vol} G_{A}$ & $\sum_{i \in N}d^i(A)$ --- volume, the sum of in-degrees \\
$E_{\max }$ & $\{ (j, i): \sup_{t\ge 0} a_t^{i,j} > 0\}$ --- maximal set of communication links \\
$x_t^i $ & state of agent $i$ at time $t$\\
$y_{t}^{i, j}$ & noisy and delayed measurement agent $i$ obtains from agent $j$ at time $t$ \\
$ w_{t}^{i, j} $ & noise in $y_{t}^{i, j}$ at time $t$ \\
$d_{t}^{i, j}$ & integer-valued delay in $y_{t}^{i, j}$ at time $t$ \\
$\bar d$ & maximal delay \\
$u_{t}^{i}$ & control actions \\
$K_{t}^{i} $ & protocol with topology $(N, E_{t})$ \\
$\bar N_t^i$ & subset of $N_t^i$ at time $t$\\
$\alpha_{t} >0$ & step sizes of the local voting protocol \\
$ b_{t}^{i, j}$ & weight parameter of the local voting protocol \\
$B_t$ & matrix of the local voting protocol \\
$\bar x_{t}$ & $[x_{t}^1; \ldots ; x_{t}^n]$ \\
$\bar u_{t}$ & $[u_{t}^1; \ldots ; u_{t}^n]$ \\
$I$ & identity matrix of size $n \times n$ \\
%$P$ & Perron matrix $P = I - {\cal L}(\alpha  A)$ \\
$\underline 1$ & vector consisting of units \\
$\bar z_1$ & left eigenvector of matrix $P$: $\bar z_1=[z^1, \ldots, z^n]$ \\
${T}(\varepsilon)$ & time to $\varepsilon$-consensus \\
$x^{\star}$ & consensus value \\
${\mathrm{E}}$ & mathematical expectation \\
$\mathrm{E}_x$ & conditional expectation under the condition $x$ \\
$\mathrm{E}_{\cal F}$ & conditional expectation with respect to $\sigma$-algebra ${\cal F}$ \\
%$p_a^{i, j}$ & probability of appearances of variable edge $(j, i)$ in graph ${\cal G}_{A_t}$ \\
$b^{i, j}$ & $\mathrm{E} b_{t}^{i, j}$ \\ %JYM: $\lim_{t \to \infty} \mathrm{E} b_{t}^{i, j}$
$p_k^{i,j}$ & probability that the delay $d_{t}^{i, j} $ equals $k$ \\
$\bar n$ &  $n(\bar d +1)$ \\
%$R(\alpha, \bar x)$: & $\left(\begin{array}{c} \cdots\\ \frac{1}{\alpha }f^i( x^i, {\alpha }s^i(\bar x)) \\ \cdots \\ \end{array}\right)$ \\
%$s^i(\bar x)$: & $\sum_{j\in  N^{i}_{\max} } a_{\max}^{i, j} (x^{j} - x^i)$ \\
$A_{\max}$ & adjacency matrix of the averaged system \\
$\tau_{t}$ & $\alpha _{0} +\alpha _{1} +\ldots +\alpha _{t-1}$ \\
$\tau _{\max}$ & $\sum_{t=0}^T \alpha_t$ \\
$ {\bar X}_{t}$ & $[\bar x_{t}, \bar x_{t-1}, \ldots, \bar x_{t-{\bar d}}]$ \\
%$U$: & $\left(\begin{array}{ccccc} {I} & {0} & {0} & {\ldots } & {0} \\ {I} & {0} & {0} & {\ldots } & {0} \\ {0} & {I} & {0} & {\ldots } & {0} \\ {\vdots } & {\vdots } & {\ddots } & {\vdots } & {\vdots } \\ {0} & {0} & {\ldots } & {I} & {0}  \end{array}\right)$ \\
%$F(\alpha_t, \bar X_{t}, \bar w_{t})$: & $\left(\begin{array}{c} \cdots \\  f^i(x^i_t, \alpha_{t} \mathop{\sum }\limits_{j\in \bar N_{t}^{i} } b_{t}^{i, j} ((x_{t-d_t^{i, j}}^j - x_t^i)+ (w_{t}^{i, j}  - w_{t}^{i, i} ))) \\ \cdots \\ {0}_{n\bar d} \end{array}\right)$ \\
%$G(\alpha, \bar Z)$: & $\left(\begin{array}{c} \cdots\\ f^i(z^i, {\alpha }s^i(\bar Z) ) \\ \cdots \\ {0}_{n\bar d} \end{array}\right)$  \\
%$s^i(\bar Z)$: & $\mathop{\sum }\limits_{j\in  N^{i} } p_a^{i,j}b^{i,j} ((\mathop{\sum }\limits_{k=0}^{\bar d}p_k^{i,j}z^{j+k n}) - z^i)$ \\
$q_{t}^{i}$ & queue length of the atomic elementary jobs of the agent $i$ at time $t$ \\
$p_{t}^{i}$ & productivity of the agent $i$ at time $t$ \\
$z_{t}^{i} $ & new job received by agent $i$ at time $t$ \\
$T_t$ & implementation time of jobs at time $t$ \\
$\frac{q_{t}^{i}}{p_{t}^{i}}$ & load of agent $i$ at time $t$ \\
$Err$ & $\sqrt{\sum _{i} \frac{(x_{t}^{i} -x^{\star})^{2} }{n} }$ --- average residual \\
$|D(t)|$ & maximum deviation from the average load on the network \\
%: &  \\
\end{supertabular}
\end{center}

%%%%%%%%%%%%%%%%%%%%%%%%%%%%%%%%%%%%%%%%%%%%%%%%%%%%%%%%%%%%%%%%%%%%%%%%%%%%%%%%
\subsection{Problem Statement}

\subsubsection{The network model}
Consider a dynamic network of $n$ agents that collaborate to solve a problem that each cannot solve alone.

The concepts of graph theory will be used to describe the network topology.
Let the dynamic network topology be modeled by a sequence of digraphs $\{(N, E_{t})\} _{t\ge 0} $, where $E_t \subset E$ changes with time. The corresponding adjacency matrices are denoted as $A_t$.
The maximal set of communication links is $ E_{\max } =\{ (j, i): \sup_{t\ge 0} a_t^{i,j} > 0\}$.

We assume that a time-varying state variable $x_t^i \in {\mathbb R}$  corresponds to each agent $i \in N$ of the graph at time $t \in [0, T]$. Its dynamics are described for the discrete time case as
\begin{equation}
\label{Nat_1}
x_{t+1}^{i} = x_{t}^{i} +f^{i} (x_{t}^{i}, u_{t}^{i}), \;t = 0, 1, 2 \ldots, T
\end{equation}
or for the continuous time case as
\begin{equation}
\label{Nat_1cont}
\dot x_{t}^{i} = f^{i} (x_{t}^{i}, u_{t}^{i}), \;t \in [0, T],
\end{equation}
with some functions $f^i(\cdot,\cdot): {\mathbb R} \times {\mathbb R} \to {\mathbb R} $, depending on states in the previous time $x_{t}^{i}$ and control actions $u_{t}^{i} \in {\mathbb R}$.

Each agent uses its own state (possibly noisy) to form its control strategy
\begin{equation}
\label{Nat_3i}
y_{t}^{i, i} =x_{t}^{i}  +w_{t}^{i, i},
\end{equation}
 and if $N^i_t \neq \emptyset$, noisy and delayed measurements of its neighbors' states
\begin{equation}
\label{Nat_3}
y_{t}^{i, j} = x_{t-d_{t}^{i, j} }^{j} + w_{t}^{i, j},\; j \in N_{t}^{i},
\end{equation}
where $w_{t}^{i, i}, w_{t}^{i, j} $ are noises, $0 \le d_{t}^{i, j} \le \bar d$ is an integer-valued delay, and $\bar d$ is a maximal delay.

%We consider the multi-agent system consisting of dynamic agents $i \in N$ with inputs $u_t^i$, outputs $y_{t}^{i, j}$ and states $x_t^i$.

If $(j, i) \in E_{t} $ then agent $i$ receives information from agent $j$ for the purposes of feedback control.

\subsubsection{The locol voting protocol}

\begin{definition} A feedback on observations
\begin{equation}
\label{Nat_2}
u_{t}^{i} =K_{t}^{i} (y_{t}^{i, j_{1} }, \ldots, y_{t}^{i, j_{m_{i} } } ),
\end{equation}
 where $\{ j_{1} ,\ldots ,j_{m_{i} } \} \in \{ i\} \bigcup \overline{N}^{i}_{t}, \; \overline{N}^{i}_{t} \subseteq  N_{t}^{i}$ is called a \textit{protocol} (\textit{control algorithm}) with topology $(N, E_{t})$. \end{definition}

In this paper, we consider the \textit{local voting protocol}:
\begin{equation}
\label{Nat_7}
u_{t}^{i} =\alpha_{t} \sum_{j\in \overline{N}^{i}_{t} } b_{t}^{i, j} (y_{t}^{i, j} - y_{t}^{i, i} ),
\end{equation}
where $\alpha_{t} >0$ are step sizes of control protocol, $ b_{t}^{i, j} > 0\; \; \forall j\in \bar N_{t}^{i}$. We set $b_{t}^{i, j} = 0$ for other pairs $i, j$ and denote $B_t = [b_{t}^{i, j}]$ as the matrix of the control protocol.

Note, that protocol~(\ref{Nat_7}) differs from a frequently used such protocol, where control step parameters $\alpha$ vary for different agents $i \in N$ (for example, $\alpha^i=1/d^i(B_t)$, see~\cite{Guo}).

\subsubsection{Consensus concepts}

In this paper, various consensus concepts will be employed, which are defined as follows.

\begin{definition}Agents $i$ and $j$ are said to \textit{agree} in a network at time $t$ if and only if $x_t^i = x_t^j$. \end{definition}

 \begin{definition} The network is said to reach a \textit{consensus} at time $t$ if  $x_t^i=x_t^j~~\forall i,j \in N, i \neq j$. \end{definition} %$n$ agents of a network

 \begin{definition} The network is said to achieve \textit{asymptotic mean square consensus} if there exists
% a variable
 $\lim_{t \to \infty} E ||x^{i}_t - x^{\star}||^2 = 0$ for all $ i \in N$. \end{definition} %$n$ agents are

 \begin{definition} The network is said to reach an \textit{average consensus} at time $t$ if all nodes' states drive to the same constant steady-state value: $x_t^i=x_t^j=c~~\forall i,j \in N, i \neq j$, where $c$ is the average of the initial states of the agents
 \begin{equation}c=\frac{1}{n}\sum_{i=1}^n x^i_0\end{equation}
 \end{definition} %$n$ agents of

Here, this value does not depend on the graph structure. The average consensus problem is important in many applications. For instance, in wireless sensor networks each agent measures some quantity (e.g., temperature, salinity content, {\it etc.}) and it is desired to determine the best estimate of the measured quantity, which is the average if all sensors have identical noise characteristics.

\begin{definition} The network is said to achieve \textit{$\varepsilon$-consensus} at time $t$ if there exists a variable $x^{\star} $ such that $||x_{t}^{i} -x^{\star} ||^{2} \leq \varepsilon$ for all $i \in N$. \end{definition} %$n$ agents are

\begin{definition} ${T}(\varepsilon)$ is called \textit{time to $\varepsilon$-consensus}, if the network achieves  $\varepsilon$-consensus for all $t \geq {T}(\varepsilon)$. \end{definition}

\begin{definition}
The network is said to achieve \textit{mean square $\varepsilon$-consensus} at time $t$ if there exists a variable $x^{\star} $ such that ${\mathrm{E}}||x_{t}^{i} - x^{\star} ||^{2} \leq \varepsilon$ for all $ i\in N$.
\end{definition}

\begin{definition}
The network is said to achieve \textit{asymptotic mean square $\varepsilon$-consensus} at time $t$ if $\mathrm{E}||x_{1}^{i} ||^{2} <\infty,\; i\in N$ and there exists a variable $x^{\star} $ such that $\mathop{\overline{\lim} }\nolimits_{t \to \infty } { \mathrm{E}}||x_{t}^{i} -x^{\star} ||^{2} \leq \varepsilon$ for all $ i\in N$.
\end{definition}

%For convenience of statistical modelling, we make the convention: $w_t^{ij}$ and $d_t^{ij}$ are defined for all $(j,i) \in E_{\max}$. If $(j,i)$ does not appear in $E_t$ so that \eqref{3} does not physically occur, we still introduce $w_t^{ij}$ and $d_t^{ij}$ and set them to zero.

%JYM: The following commented part has been moved up.
%Consider the \textit{local voting protocol}:
%\begin{equation}\label{Nat_7}
%u_{t}^{i} =\alpha_{t} \sum_{j\in \overline{N}^{i}_{t} } b_{t}^{i, j} (y_{t}^{i, j} - y_{t}^{i, i} ),
%\end{equation}
%where $\alpha_{t} >0$ are step sizes of control protocol, $ b_{t}^{i, j} > 0\; \; \forall j\in \bar N_{t}^{i}$. We set $b_{t}^{i, j} = 0$ for other pairs $i, j$ and denote $B_t = [b_{t}^{i, j}]$ as the matrix of the control protocol.

%Note, that protocol~(\ref{Nat_7}) differs from the other frequently using protocol, where control step parameters $\alpha$ vary for different agents $i \in N$ (for example, $\alpha^i=1/d^i(B_t)$, see~\cite{Guo}).

%%%%%%%%%%%%%%%%%%%%%%%%%%%%%%%%%%%%%%%%%%%%%%%%%%%%%%%%%%%%%%%%%%%%%%%%%%%%%%%%
\subsection{Preliminary Results}

Consider the particular case of dynamic systems on graphs when the second term in (\ref{Nat_1}) has a simple form: $f^i(x_t^i, u_t^i)=u_t^i$, for all agents $i$, and all observations are made without noise and delays: $y_t^{i, j} = x_t^j, \; j \in \{i\}\cup N_t^i.$

Denote $\bar x_{t}=[x_{t}^1; \ldots ; x_{t}^n]$ and $\bar u_{t}=[u_{t}^1; \ldots ; u_{t}^n]$  column vectors obtained by the vertical concatenation of $n$ corresponding variables.
Control protocol~(\ref{Nat_7}) can be rewritten in a matrix form:
\begin{equation}
\label{Nat_7m}
\bar u_{t} = (\alpha_t B_t - D(\alpha_t B_t) ) \bar x_{t} = -  {\cal L}(\alpha_t B_t) \bar x_{t}.
\end{equation}

The dynamics~(\ref{Nat_1}) for the discrete time case is described by:
\begin{equation}
\label{Nat_1m}
\bar x_{t+1} = \bar x_{t} + \bar u_{t},\;t=0, 1, 2, \ldots, T,
\end{equation}
and for the continuous-time case by:
\begin{equation}
\label{Nat_1mdiff}
\dot{\bar{ x}}_{t} = \bar u_{t},\;t \in [0, T].
\end{equation}

%При этом замкнутая система в дискретном времени принимает вид:
With (\ref{Nat_7m}), the dynamics of the closed-loop system for the discrete time case takes the form:
\begin{equation}
\label{Nat_1z}
\bar x_{t+1} = (I -  {\cal L}(\alpha_t B_t)) \bar x_{t},\;t=0, 1, 2, \ldots, T,
\end{equation}
where $I$ is matrix of size $n \times n$ of ones and zeros on the diagonal,
and for the continuous time case the dynamics takes the form
\begin{equation}
\label{Nat_1zdiff}
\dot{\bar{ x}}_{t}  =  -  {\cal L}(\alpha_t B_t) \bar x_{t},\;t \in [0, T].
\end{equation}

We will show that the control protocol~(\ref{Nat_7}) with $\alpha_t = \alpha$ and $B_t =  A$  provides consensus asymptotically for both discrete and continuous-time models.  Similar results can be found in \cite{Olfati, Lewis2011}.

\subsubsection{The discrete-time case}

\begin{Lemma}\label{Nat_DiscCons} %\cite{Amelina12SNPD}
If the graph ${\cal G}_A$ has a spanning tree, and for the  control protocol~\eqref{Nat_7}, we have  parameters  $B_t =  A$ and $\alpha_t = \alpha$ such that the following condition is satisfied
\begin{equation}
\label{Nat_dmax} \alpha < \frac{1}{d_{\max}},
\end{equation}
 then the control protocol~\eqref{Nat_7} provides asymptotic consensus for the discrete system~(\ref{Nat_1m}) and its value $x^{\star}$ is given by~\eqref{Nat_DiscConsValue}.
 \end{Lemma}

 \begin{proof}
 Indeed, for the discrete case the equation~(\ref{Nat_1z}) turns into
 \begin{equation}
 \label{Nat_1z13}
 \bar x_{t+1} = (I -  {\cal L}(\alpha  A)) \bar x_{t} \equiv P \bar x_{t},
 \end{equation}
 where the Perron matrix $P = I - {\cal L}(\alpha  A)$ has one simple eigenvalue equal to one and all others are inside the unit circle if the condition~\eqref{Nat_dmax} is satisfied.
Since the sum of row elements of the Laplacian ${\cal L}$ equals to zero, the sum of row elements of matrix $P$ equals to one, i.e. vector $\underline 1$ consisting of units is a right eigenvector of $P$ corresponding to the unit eigenvalue. The unit eigenvalue is simple if the graph has a spanning tree. All other eigenvalues are inside the unit circle. Let $\bar z_1=[z^1, \ldots, z^n]$ denote the left eigenvector of matrix $P$ which is orthogonal to $\underline 1$. Consequently, if the graph has a spanning tree then in the limit of $t \to \infty$ we get
 \begin{equation}
\bar x_t \to {\underline 1} (\bar z_1^{\rm T} \bar x_0),
 \end{equation}
i.e. an asymptotic consensus is reached. The consensus value $x^{\star}$ equals to the normalized linear combination of initial states with weights equal to elements of the left eigenvector of matrix~$P$
 \begin{equation}\label{Nat_DiscConsValue}
x^{\star} = \frac{\bar z_1^{\rm T} \bar x_0}{\bar z_1^{\rm T} {\underline 1}} = \frac{ \sum_{i=1}^n z^i x^i_0}{\sum_{i=1}^n z^i}.
 \end{equation}
This value depends on the graph topology and, consequently, on connection links between agents.
 \end{proof}

% Similar result was presented in \cite{Olfati, Lewis2011}. %JYM: It has been moved up.

%JYM: The following commented part has been moved up.
% \begin{definition} $n$ agents of a network are said to reach an \textit{average consensus} at time $t$ if all states drive to the same constant steady-state value: $x_t^i=x_t^j=c~~\forall i,j \in N, i \neq j$, where $c$ is the average of the initial states of the agents
% \begin{equation}c=\frac{1}{n}\sum_{i=1}^n x^i_0\end{equation}
% \end{definition}

%This value does not depend on the graph structure. The average consensus problem is important is many applications. For instance, in wireless sensor networks each agent measures some quantity (e.g., temperature, salinity content, {\it etc.}) and it is desired to determine the best estimate of the measured quantity, which is the average if all sensors have identical noise characteristics.

\begin{Lemma}\label{Nat_AverConsSensor} If the graph is balanced then the sums of the rows of the Laplacian ${\cal L}$ is equal to the sum of the corresponding columns, and this property is transferred to the matrix $P$ then $\bar z_1= c {\underline 1}$,
and the consensus value equals to the initial values average
 \begin{equation}
 \nonumber
x^{\star} = \frac{1}{n}  \sum_{i=1}^n x^i_0
 \end{equation}
and does not depend on the topology of the graph.
 \end{Lemma}

 \begin{proof}
The conclusion of Lemma \ref{Nat_AverConsSensor} follows directly from Lemma~\ref{Nat_DiscCons}, since in the balanced case, $z^i$ from~(\ref{Nat_DiscConsValue}) are equal to 1, i.e. the left and right eigenvectors corresponding to the zero eigenvalue are equal.
 \end{proof}

\subsubsection{The continuous-time case}

\begin{Lemma} \cite{Lewis2011} \label{Nat_ContCons} %\cite{Amelina12SNPD}
If the graph ${\cal G}_A$ has a spanning tree then the control protocol~\eqref{Nat_7} with $\alpha_t = \alpha$ and $B_t =  A$ provides an asymptotic consensus for the continuous-time system~(\ref{Nat_1mdiff}) and its value $x^{\star}$ is given by
\begin{equation}
\label{Nat_ContConsValue}
x^{\star} = \frac{1}{\sqrt{n}}\sum_{i=1}^n \bar z_1^i x^i_0
\end{equation}
with vector of initial data $\bar x_0$ and the orthonormal first left eigenvector $\bar z_1$ of the matrix ${\cal L}$.
 \end{Lemma}
  \begin{proof}
For the continuous-time case we have
\begin{equation}
\label{NL1}
\dot {\bar x} = -  {\cal L} {\bar x}.
\end{equation}

Let $\bar z_1, \bar z_2, \ldots, \bar z_n$ and $\bar r_1= \frac{1}{\sqrt{n}}\underline 1, \bar r_2,  \ldots, \bar r_n$ be left and right orthonormal eigenvectors of the matrix ${\cal L}$ corresponding to its ordered eigenvalues $\lambda_1, \ldots, \lambda_n$.
If the graph has a spanning tree then $\lambda_1=0$  is a simple eigenvalue and all other eigenvalues of ${\cal L}$ are in the open right half of complex plane. Thus, the system~(\ref{NL1}) is partially stable with one pole at the origin and the rest are in the open left half plane.

For the first left eigenvector $\bar z_1=[\bar z^1, \ldots, \bar z^n]$ of matrix ${\cal L}$ we have
\begin{equation}
\frac{d}{dt}(\bar z_1^{\rm T} \bar x_t) = \bar z_1^{\rm T} \dot {\bar x}_t = - \bar z_1^{\rm T}  {\cal L} \bar x_t = 0,
\end{equation}
i.e. $\tilde x \equiv \bar z_1^T \bar x_t = \sum_{i=1}^n  z_1^i x_t^i$ is invariant, that is constant and independent of the states of agents. Thus, $\sum_{i=1}^n \bar z_1^i x^i_0=\sum_{i=1}^n \bar z_1^i x^i_t, \forall t$.

We apply the modal expansion and rewrite the state vector in terms of eigenvalues and eigenvectors of the matrix ${\cal L}$. If all the eigenvalues of ${\cal L}$ are simple (in fact, it is only important that $\lambda_1$ is simple), then
\begin{equation}
\label{NL3}
\bar x_t = e^{ -  {\cal L} t} \bar x_0 = \sum_{j=1}^n \bar r_j e^{-\lambda_j t} \bar z_j^{\rm T} \bar x_0 = \sum_{j=2}^{n}  (\bar z_j^{\rm T} \bar x_0) e^{-  \lambda_j t} \bar r_j + \frac{\tilde x}{\sqrt{n}} \underline 1.
\end{equation}
In the limit of $t \rightarrow \infty$ we get $x_t \rightarrow \frac{\tilde x}{\sqrt{n}} \underline 1$ or $x^i_t \rightarrow x^{\star}=\frac{\tilde x}{\sqrt{n}}, \; \forall i \in N$, i.e. an asymptotic consensus is reached.

 \end{proof}

%Similar result was presented in~\cite{Lewis2011}.

%%%%%%%%%%%%%%%%%%%%%%%%%%%%%%%%%%%%%%%%%%%%%%%%%%%%%%%%%%%%%%%%%%%%%%%%%%%%%%%%%
%\subsection{Approximate Consensus}

%On a finite time interval we can talk
In the continuous-time case, we focus on the problem of reaching an approximate $\varepsilon$-consensus ($\varepsilon > 0$).

%JYM: The following commented part has been moved up.
% \begin{definition} $n$ agents are said to achieve \textit{$\varepsilon$-consensus} at time $t$ if there exists a variable $x^{\star} $ such that $||x_{t}^{i} -x^{\star} ||^{2} \leq \varepsilon$ for all $i \in N$. \end{definition}

% \begin{definition} ${T}(\varepsilon)$ is called \textit{time to $\varepsilon$-consensus}, if $n$ agents achieve  $\varepsilon$-consensus for all $t \geq {T}(\varepsilon)$. \end{definition}

%Здесь $\lambda_1 = 0$ и берем $\bar v_1 = \underline 1$. Определим левый собственный вектор $\bar z_1=[z^1, \ldots, z^n]$ нормированный так, что $\bar z_i^T \bar v_i = 1$, то есть $\sum_i z^i=1$. Тогда получается:
%$$
%x^t \rightarrow \bar v_2 e^{-\lambda_2t} \bar z_2^T x_0 + \bar v_1 e^{-\lambda_1t} \bar z_1^T x_0 = \bar v_2 %e^{-\lambda_2t} \bar z_2^T x_0 + \underline 1 \sum_{i=1}^n z^i x_0^i
%$$
%Последний член в этом уравнении определяет значение консенсуса~(\ref{NL4}) (отметим, что в~(\ref{NL4}) левый собственный вектор не нормирован). Первое слагаемое показывает, что значение консенсуса достигается с постоянным временем \textbf{(is reached with a time constant given by)} $\tau=1/\lambda_2$.

\begin{Lemma}\label{Nat_ContConsTime} %\cite{Amelina12SNPD}
If the graph ${\cal G}_A$ has a spanning tree, then the control protocol~\eqref{Nat_7} with $\alpha_t = \alpha$ and $B_t =  A$ provides $\varepsilon$-consensus for the continuous-time system~(\ref{Nat_1mdiff}) for any $t \geq { T}(\varepsilon)$, where ${ T}(\varepsilon)$ is defined by:
\begin{equation}
\label{NL_Teps}
{ T}(\varepsilon) = \frac{1}{2 Re(\lambda_2)} \ln\left(\frac{(n-1)||x_0-x^{\star}\underline 1||^2}{\varepsilon}\right),
\end{equation}
and the consensus value  $x^{\star}$ is given by the formula~\eqref{Nat_ContConsValue}.
 \end{Lemma}

 \begin{proof}
From (\ref{NL3}) by evaluating the square of the norm of the first term we can obtain
\begin{equation}||\bar x_t-x^{\star}\underline 1||^2 = || \sum_{j=2}^{n}  (\bar z_j^{\rm T} \bar x_0) e^{-  \lambda_j t} \bar r_j||^2 = \end{equation}
$$= || \sum_{j=2}^{n}  (\bar z_j^{\rm T} (\bar x_0-x^{\star})) e^{-  \lambda_j t} \bar r_j||^2 \leq (n-1)e^{- 2  Re(\lambda_2) t} ||\bar x_0-x^{\star}\underline 1||^2.
$$ From here we have the expression~(\ref{NL_Teps}) for the time to $\varepsilon$-consensus in the system~(\ref{NL1}).
 \end{proof}

\vspace{3mm}
Here we highlight that, in contrast to the earlier results using $||x_0||^2$ in \cite{Lewis2011}, we have considered $||x_0-x^{\star}\underline 1||^2$ instead of $||x_0||^2$ inside the argument of $\ln$-function.

%A similar estimate for the time to $\varepsilon$-consensus can be obtained for the discrete system~(\ref{Nat_1z13}).

%%%%%%%%%%%%%%%%%%%%%%%%%%%%%%%%%%%%%%%%%%%%%%%%%%%%%%%%%%%%%%%%%%%%%%%%%%%%%%%%
\section{Main Results} \label{sec-main}

In this section, we present the main results of this paper. All proofs are included in the Appendix.

%%%%%%%%%%%%%%%%%%%%%%%%%%%%%%%%%%%%%%%%%%%%%%%%%%%%%%%%%%%%%%%%%%%%%%%%%%%%%%%%
\subsection{Main Assumptions}

Let $({\Omega},{\cal F},{P})$ be the underlying probability space corresponding to the sample space, the collection of all events, and the probability measure respectively.
%Let ${\mathrm{E}}$ be the symbol for the mathematical expectation and $\mathrm{E}_x$ be the conditional expectation under the condition $x$.

For the remaining article, we assume that the following conditions are satisfied:

\vspace{2mm}

{\bf A1}. $\forall i \in N $ functions $f^i(x, u)$ are Lipschitz in $x$ and $u$:
$|f^i(x, u)-f^i(x', u')| \leq L_1(L_x|x-x'| + |u-u'|)$,
%the growth rate is bounded: $|f^i(x, u)|^2 \leq L_{2}(L_c+L_x|x|^2+|u|^2)$,
and for any fixed $x$ the function $f^i(x, \cdot)$ is such that $\mathrm{E}_x f^i(x, u) = f^i(x, \mathrm{E}_x\, u)$.
Note that, following from this Lipschitz condition, the growth rate is bounded:
$|f^i(x, u)|^2 \leq L_{2}(L_c+L_x|x|^2+|u|^2)$.

\vspace{2mm}

{\bf A2}.  {\bf a)} $\forall i \in N, j \in N^{i}_{\max}$ the noises $w_{t}^{i, j}$ are centered, independent and have bounded variance $E (w_{t}^{i, j})^2\leq \sigma_w^2$.

%\item
%{\bf A3}.
{\bf b)}
$\forall i \in N, j \in N^{i}_{\max}$ appearances of variable edges $(j, i)$ in graph ${\cal G}_{A_t}$ are independent random events.
 %with probability $p_a^{i, j}$ (i.e., matrices $A_t$ are independent, identically distributed random matrices).

%\item
%{\bf A4}.
{\bf c)}
$\forall i \in N, j \in N^{i}_{\max}$ weights $b_{t}^{i, j}$ in the control protocol are independent random variables with %such that
%$\underline b \leq b_{t}^{i, j} \leq \bar b$ with probability 1, and there exist limits
$b^{i, j} = \mathrm{E} b_{t}^{i, j},~\sigma_b^{i, j}=E(b_{t}^{i, j}-b^{i, j})^2 < \infty$.

{\bf d)} $\forall i \in N, j \in N^{i}  $ there exists a finite quantity $\bar d \in \mathbb N$: $d_{t}^{i, j} \leq \bar d$ with probability 1 and integer-valued delays $d_{t}^{i, j} $ are  independent, identically distributed random variables taking values $k=0, \ldots, \bar d$ with probabilities $p_k^{i,j}$.

%{\bf e)}
More over, all these random variables and matrices are mutually independent.

\vspace{2mm}

The next assumption is for a matrix $A_{\max}$ constructed as follows. Specifically, if $\bar d>0$, we add new ``fictitious'' agents whose states at time $t$ equal to the corresponding states of the ``real'' agents at the previous $\bar d$ time: $t-1, t-2, \ldots, t-\bar d$. Then, $A_{\max}$ is a matrix  of size $\bar n \times \bar n $, where $\bar n = n \times (\bar d +1)$, with  %JYM
%Specifically, $A_{\max}$ is a matrix  of size $\bar n \times \bar n $ and consider the network topology with $\bar n = n(\bar d +1)$ agents as:
\begin{equation} a_{\max}^{i, j} =   p_{j \div \bar d}^{i, j \, {\rm mod}\, \bar d} \, b^{i, j \, {\rm mod}\, \bar d}, \;i\in N, \; j= 1, 2, \ldots, \bar n,\end{equation}
$$ a_{\max}^{i, j} =   0, \;i = n+1, n+ 2, \ldots, \bar n, \; j= 1, 2, \ldots, \bar n.$$
Here, the operation $\mod$ is a remainder of division, and $\div$ is a division without remainder.

Note that if $\bar d = 0$, this definition of network topology (of matrix $A_{\max}$ of size $n \times n$) is reduced to
\begin{equation}
a_{\max}^{i, j} = b^{i, j}, \;i\in N,\; j \in N.
\end{equation}

Also note that we have defined a matrix $ A_{\max} $ in such a way that $E_{\bar x_t} \bar u_t = -\alpha_t {\cal L} (A_{\max}) \bar x_t$. We assume that the following condition is satisfied for this network topology matrix:

%\begin{description}
%\item
{\bf A3}. Graph $ (N, {E}_{\max})$
%, определяемый матрицей смежности $A_{\max}$, %с элементами
%является сильно связным.
has a spanning tree, and for any edge $(j, i) \in {E}_{\max}$ among the elements $a_{\max}^{i, j}, a_{\max}^{i, j+n}, \ldots, a_{\max}^{i, j+\bar d n} $ of the matrix $A_{\max}$, there exists at least one non-zero.
%\end{description}

%JYM: They are moved above together with other consensus concepts.
%\begin{definition}
% $n$ agents are said to achieve \textit{mean square $\varepsilon$-consensus} at time $t$ if there exists a variable $x^{\star} $ such that ${\mathrm{E}}||x_{t}^{i} - x^{\star} ||^{2} \leq \varepsilon$ for all $ i\in N$.
%\end{definition}

%\begin{definition}
% $n$ agents are said to achieve \textit{asymptotic mean square $\varepsilon$-consensus} at time $t$ if $\mathrm{E}||x_{1}^{i} ||^{2} <\infty,\; i\in N$ and there exists a variable $x^{\star} $ such that $\mathop{\overline{\lim} }\nolimits_{t \to \infty } { \mathrm{E}}||x_{t}^{i} -x^{\star} ||^{2} \leq \varepsilon$ for all $ i\in N$.
%\end{definition}

%%%%%%%%%%%%%%%%%%%%%%%%%%%%%%%%%%%%%%%%%%%%%%%%%%%%%%%%%%%%%%%%%%%%%%%%%%%%%%%%

\subsection{The Case without Delay in Measurement}
%\section{Analysis of the Closed Loop System Dynamics}

We first consider the case where there is no delay in measurement, i.e. $\bar d = 0$. % \cite{Amelina12, Amelina12SNPD}.

%Denote $\bar x_{t}=[x_{t}^1; \ldots ; x_{t}^n]$.
Rewrite the dynamics of the agents in the vector-matrix form:
\begin{equation}
\label{Nat_821}
\bar x_{t+1} = \bar x_{t} +  F(\alpha_{t}, \bar  x_{t}, \bar w_{t}),\quad \quad
\end{equation}
where
$F(\alpha_{t}, \bar x_{t}, \bar w_{t})$ is the vector of dimension $n$:
$$F(\alpha_{t}, \bar x_{t}, \bar w_{t})=$$
\begin{equation}
\label{Nat_9B21}
=\left(\begin{array}{c} \cdots \\  f^i(x_{t}^i, \alpha_{t}\mathop{\sum }\limits_{j\in \bar N_{t}^{i} } b_{t}^{i,j} ((x_t^j - x_t^i)+ (w_{t}^{i, j}  - w_{t}^{i, i} ))) \\ \cdots  \end{array}\right).
\end{equation}

To analyze the stochastic system behavior at the particular choice of the coefficients (parameters), in the control protocol, it is common to use the {\it method of averaged models} \cite{DF74ARC}, (also called ODE approach \cite{Ljung77}, or Derevitskii-Fradkov-Ljung (DFL)-scheme \cite{Gerencser06}), which we also adopt in this paper.

Specifically in our use, the method of averaged models consists on the approximate replacement of the initial stochastic difference equation \eqref{Nat_821} by an ordinary differential equation: %JYM
\begin{equation}
\label{e2}
\frac{ d \bar x}{d \tau}=R(\alpha, \bar x),
\end{equation}
where
\begin{equation}
\label{Nat_9D21}
R(\alpha, \bar x) = R\left(\alpha, \begin{array}{c} x^1\\  \vdots \\ x^{n} \end{array}\right) =\left(\begin{array}{c} \cdots\\ \frac{1}{\alpha }f^i( x^i, {\alpha }s^i(\bar x)) \\ \cdots \\ \end{array}\right),
\end{equation}
$$
s^i(\bar x)=\sum_{j\in  N^{i}_{\max} } a_{\max}^{i, j} (x^{j} - x^i) = - d^i(A_{\max})x^i + \sum_{j=1}^{n} a_{\max}^{i, j} x^j, \,i \in N.
$$
where $A_{\max}$ is the adjacency matrix whose construction is introduced in the previous subsection.

Note, that if the last part of the condition A1 is not satisfied, then instead of \eqref{Nat_9D21} one can use the following definition
\begin{equation}
\label{Nat_9D21-2}
R(\alpha, \bar x) = \frac{1}{\alpha}  \mathrm{E}_x  F(\alpha_{t}, \bar x_{t}, \bar w_{t}).
\end{equation}

%Matrix $A_{\max}$ is the adjacency matrix for the averaged system.

%The trajectories of $\{ \bar x_{t} \} $ from~\eqref{Nat_821}-\eqref{Nat_9B21} and of $\{{\bar x}(\tau _{t} )\} $ from~\eqref{e2}-\eqref{Nat_9D21} are close in a finite time interval~\cite{DF74ARC}.
According to \cite{DF74ARC}, the trajectories of $\{ \bar x_{t} \} $ from~\eqref{Nat_821}-\eqref{Nat_9B21} and of $\{{\bar x}(\tau _{t} )\} $ from~\eqref{e2}-\eqref{Nat_9D21} are close in a finite time interval.
Here and below let $\tau_{t} =\alpha _{0} +\alpha _{1} +\ldots +\alpha _{t-1}$. %In particular, if $\alpha_t = \alpha = const \; \forall t$, then $\tau_{t} =T \alpha$.

In the following theorem the upper bounds for mean square distance between the initial system and its averaged continuous model will be given.

\vspace{3mm}
\begin{Theorem}\label{Nat_L11}
If conditions {\bf A1}, {\bf A2a--c} are satisfied, $\forall i \in N $ function $f^i(x, u)$ is smooth in $u$, $f^i(x,0)=0$ for any  $x$, and $0<\alpha_t \leq \bar \alpha$,
{\bf then} there exists $\tilde \alpha$ such that for $\bar \alpha < \tilde \alpha$ the following inequality  holds:
 \begin{equation}\mathrm{E}\mathop{\max }\limits_{0\le \tau _{t} \le \tau _{\max} } {\rm ||} \bar x_{t} - {\bar x}(\tau _{t} ){\rm ||}^{2} \le C_{1} e^{C_{2} \tau _{\max} } \bar \alpha, \end{equation} where
$C_{1} >0$, $C_{2} >0$ and $\bar \alpha >0$ are some constants.
\end{Theorem}
%Proof is in the Appendix.

We return to the problem of achieving consensus. Assume that, in the averaged continuous model~\eqref{e2}-\eqref{Nat_9D21}, the $\frac{\varepsilon}{4}$-consensus is reached over time, i.e. all components of the vector $\bar x(\tau)$ become close to some common value $x^{\star}$ for all $i \in N$. Then, we have the following result.
%Условия достижения $\varepsilon$-консенсуса для траекторий $\{ \bar x_{t} \} $  из~\eqref{Nat_821}-\eqref{Nat_9B21} даны в следующей теореме.

\vspace{2mm}
\begin{Theorem}\label{Nat_T11}
Let the conditions {\bf A1}, {\bf A2a--c} be satisfied, $\forall i \in N $ functions $f^i(x, u)$ are smooth by $u$, $f^i(x,0)=0$ for any $x$, $0<\alpha_t \leq \bar \alpha$, for the continuous model~\eqref{e2}-\eqref{Nat_9D21} the $\frac{\varepsilon}{4}$-consensus is achieved for time ${\cal T}(\frac{\varepsilon}{4})$, consensus protocol parameters $\{\alpha_t\}$ are chosen so that $\tau _{\max} = \sum_{t=0}^T \alpha_t > {\cal T}(\frac{\varepsilon}{4})$ and for some constants $C_{1}, C_{2}$ the following inequality holds
\begin{equation}
C_{1} e^{C_{2} \tau _{\max} } \max_{\alpha_t: \tau_t \leq \tau _{\max}} \alpha_t \leq \frac{\varepsilon}{4},
\end{equation}
 \textbf{then} the mean square $\varepsilon$-consensus in the stochastic discrete system~\eqref{Nat_821}-\eqref{Nat_9B21} at time $t:\;{\cal T}(\frac{\varepsilon}{4}) \leq t \leq \tau _{\max}$  is achieved.
\end{Theorem}
%Proof is in the Appendix.

\vspace{2mm}

Consider an important special case where $\forall i \in N \; f^i(x, u) = u$. In this case, the time to $\frac{\varepsilon}{4}$-consensus in the averaged continuous model~\eqref{e2}-\eqref{Nat_9D21} can be obtained from Lemma~\ref{Nat_ContConsTime}:
\begin{equation}
\label{T33-1}
{\cal T}\left(\frac{\varepsilon}{4}\right) = \frac{1}{2 Re(\lambda_2)} \ln\left(\frac{4(n-1)||\bar x_0-x^{\star}\underline 1||^2}{{\varepsilon}}\right).
\end{equation}
Then, based on Theorem~\ref{Nat_T11}, the following important consequence is obtained.

\begin{Corollary} \label{Nat_T22} %JYM: Collorary?
If $f^i(x,u)=u$ for any $i \in N $, conditions {\bf A2a--c}, {\bf A3} are satisfied, $\forall i \in N $ functions $f^i(x, u)$ are smooth in $u$, $f^i(x,0)=0$ for any $x$,
{\bf then} for any arbitrarily small positive number $ \varepsilon >0$ for any $\tau_{\max}>{\cal T}(\frac{\varepsilon}{4})$ denoted in \eqref{T33-1}, when $\alpha_t$ is selected as sufficiently small
\begin{equation}\max_{\alpha_t: \tau_t \leq \tau _{\max}} \alpha_t \leq \frac{\varepsilon}{4C_1 e^{C_2 \tau_{\max}}}\end{equation}
at time $t:\;{\cal T}(\frac{\varepsilon}{4}) \leq t \leq \tau _{\max}$ in the stochastic discrete system~\eqref{Nat_821}-\eqref{Nat_9B21}, the mean square $\varepsilon$-consensus for $n$ agents is achieved,
where $C_1, C_2$ are some constants and $\lambda_2$ is the closest to the imaginary axis eigenvalue of matrix ${\cal L}$ with nonzero real part. %$C_1, C_2, \bar \alpha$ %JYM
\end{Corollary}
%Proof is in the Appendix.

%%%%%%%%%%%%%%%%%%%%%%%%%%%%%%%%%%%%%%%%%%%%%%%%%%%%%%%%%%%%%%%%%%%%%%%%%%%%%%%%

%\subsection{General Case. Analysis of the Closed Loop System Dynamics}
\subsection{The General Case with Delay in Measurement}

We now consider the general case, where $\bar d \geq 0$. % \cite{Amelina12AiT}.

Let $\bar x_t \equiv 0$ for $- {\bar d} \leq t < 0$, and denote $ {\bar X}_{t} \in {\mathbb R}^{n{\bar d}} $ as the extended state vector $ {\bar X}_{t} =[\bar x_{t}, \tilde x_{t-1}, \ldots, \tilde x_{t-{\bar d}}]$, where $\tilde x_{t-k}$ is a vector consisting of such $x_{t-k}^i$ that $\exists j \in N^i ~\exists k' \geq k: p_{k'}^{i,j} >0$, i.e. this is a value with positive probability involved in the formation of at least one of the controls.
%это значение с положительной вероятностью участвует в формировании хотя бы одного из управляющих воздействий. В дальнейшем для простоты будем считать, что так введенный расширенный вектор  состояний равен
To simplify, we assume that so introduced an extended state vector is
$ {\bar X}_{t} =[\bar x_{t}, \bar x_{t-1}, \ldots, \bar x_{t-{\bar d}}]$, i.e. it includes all the components with all kinds of delays not exceeding
%т.~е. в него входят все компоненты со всевозможными задержками, не превосходящими
$\bar d$.

Rewrite the dynamics of the agents in vector-matrix form:
\begin{equation}
\label{Nat_8}
\bar X_{t+1} = U  \bar X_{t} + F(\alpha_t, \bar  X_{t}, \bar w_{t}),\quad \quad
\end{equation}
where
%$0$ --- матрица из нулей,
 $U$ is the following matrix of size $\bar n \times \bar n$:
 \begin{equation}
\label{Nat_9}
U = \left(\begin{array}{ccccc} {I} & {0} & {0} & {\ldots } & {0} \\ {I} & {0} & {0} & {\ldots } & {0} \\ {0} & {I} & {0} & {\ldots } & {0} \\ {\vdots } & {\vdots } & {\ddots } & {\vdots } & {\vdots } \\ {0} & {0} & {\ldots } & {I} & {0}  \end{array}\right),
\end{equation}
 where $I$ is  the identity matrix of size $n \times n$, and $F(\alpha_t, \bar X_{t}, \bar w_{t}): {\mathbb R} \times {\mathbb R}^{\bar n} \times {\mathbb R}^{n^2} \to {\mathbb R}^{\bar n}$ ---
 vector function of the arguments:
% вектор-функция соответствующих аргументов:
$$F(\alpha_t, \bar X_{t}, \bar w_{t})=$$
\begin{equation}
\label{Nat_9B}
=\left(\begin{array}{c} \cdots \\  f^i(x^i_t, \alpha_{t} \mathop{\sum }\limits_{j\in \bar N_{t}^{i} } b_{t}^{i, j} ((x_{t-d_t^{i, j}}^j - x_t^i)+ (w_{t}^{i, j}  - w_{t}^{i, i} ))) \\ \cdots \\ {0}_{n\bar d} \end{array}\right),
\end{equation}
containing non-zero components only on the first $n$ places.
%содержащая ненулевые компоненты только на первых $n$ местах.

Consider the averaged discrete model corresponding to~\eqref{Nat_8}:
%Рассмотрим соответствующую~\eqref{Nat_8} усредненную дискретную модель
\begin{equation}
\label{Nat_10_Zt}
\bar Z_{t+1} = U  \bar Z_{t} + G(\alpha_{t},\bar Z_{t}), \; \bar Z_0=\bar X_{0},
\end{equation}
where
 \begin{equation}
\label{Nat_9D}
G(\alpha, \bar Z)=G\left(\alpha, \begin{array}{c} z^1\\  \vdots \\ z^{n(\bar d +1)} \end{array}\right) = \left(\begin{array}{c} \cdots\\ f^i(z^i, {\alpha }s^i(\bar Z) ) \\ \cdots \\ {0}_{n\bar d} \end{array}\right),
\end{equation}
 \begin{equation}
s^i(\bar Z)=\mathop{\sum }\limits_{j\in  N^{i} } p_a^{i,j}b^{i,j} ((\mathop{\sum }\limits_{k=0}^{\bar d}p_k^{i,j}z^{j+k n}) - z^i) =
\end{equation}
$$
= - d^i(A_{\max})z^i + \sum_{j=1}^{\bar n} a_{\max}^{i, j} z^j, \,i \in N.
$$

It turns out that
%Оказывается, что
%В главе 1 и работах \cite{DF74ARC, DF1981} показано, что при небольших дополнительных предположениях
% траектории решения исходной системы
 the trajectory of solutions of the initial system $\{\bar X_{t} \} $ from~\eqref{Nat_8} at time $t$ is close in the mean square sense to the average trajectory of the discrete system~\eqref{Nat_10_Zt}.
 %близки в среднеквадратичном смысле к траекториям усредненной дискретной системы~\eqref{Nat_10_Zt}.
 %точке траектории  $\{{X}(\tau _{t} )\} $ из~\eqref{Nat_10}, взятой при $\tau_{t} =2^{\bar d} (\alpha _{0} +\alpha _{1} +\ldots +\alpha _{t-1} )$.

%Идеи доказательства следующей теореме близки результатам из \cite{DF74ARC, DF1981}.

In the following theorem the upper bounds for mean square distance between the initial system and its averaged discrete model will be given.

\begin{Theorem}\label{Nat_T1}
%If conditions {\bf A1}, {\bf A2} are satisfied, {\bf then} there exists $\tilde \alpha$ such that for $0<\alpha_t \leq \bar \alpha < \tilde \alpha$, where $\bar \alpha$ denotes an upper bound on $\alpha_t$,
If conditions {\bf A1}, {\bf A2} are satisfied, $0<\alpha_t \leq \bar \alpha$, {\bf then} there exists $\tilde \alpha$ such that for $ \bar \alpha < \tilde \alpha$,
%JYM: See Natalia email 15.05.2013: "If conditions {\bf A1}, {\bf A2} are satisfied, $0<\alpha_t \leq \bar \alpha$ {\bf then} there exists $\tilde \alpha$ such that for $ \bar \alpha < \tilde \alpha$ the following inequality holds <...>."
the following inequality holds:
 \begin{equation}
\mathrm{E} \mathop{\max }\limits_{0\le {t} \le T } || \bar X_{t} -\bar Z_{t} ||^2  \leq c_{1}\tau_{T} e^{c_2 \tau_{T}^2 }\bar \alpha ,  %E\mathop{\max }\limits_{0\le \tau _{t} \le \tau _{max} } {\rm ||} \bar X_{t} - {X}(\tau _{t} ){\rm ||}^{2} \le C_{1} e^{C_{2} \tau _{max} } \bar \alpha,
 \end{equation}
 where $\tau_{T} =2^{\bar d} (\alpha _{0} +\alpha _{1} +\ldots +\alpha _{T-1} ),\;
c_{1}, c_2 >0$ are some constants:
\begin{equation}
c_1 = 8 n  \left(\tilde c  + \hat c(\frac{n  L_2 L_c + \bar \alpha^2 \tilde c}{c_3}+||\bar X_0||^2) e^{T\ln (c_3+1)}\right),
\end{equation}
 \begin{equation}
\tilde c = n L_1^2 \sigma_w^2 \bar b , \;c_2 =  2^{1-\bar d} L_1^2 (\frac{L_x}{\underline \alpha}+2\bar \alpha^2||{\cal L}(A_{\max})||_2^2),
\end{equation}
 \begin{equation}
c_3 =  \tilde d+L_x(2^{1+\tilde d/2}L_1 +L_2) + \bar \alpha c',\; \hat c = 2 L_1^2 n \bar b,
\end{equation}
 \begin{equation}
c' =2^{1+\tilde d/2}L_1||{\cal L}(A_{\max})||_2 + {\bar \alpha}(L_2||{\cal L}(A_{\max})||_2^2 +\hat c),
\end{equation}
 \begin{equation}
\bar b = \max_i \sum_{j=1}^n (b^{i, j})^2 + \sigma_b^{i, j},
\end{equation}
$\underline \alpha =\mathop{\min }\limits_{1\le t\le T} \alpha _{t},\;
\tilde d = 0$ if $\bar d =0$, or $\tilde d = 1$ if $\bar d >0$.
\end{Theorem}
%{\it Д~о~к~а~з~а~т~е~л~ь~с~т~в~о:}

Note that in the case without delay in measurement ($\bar d = 0$) and if $L_x=0$, then constant $c_3$ which is defined in Theorem~\ref{Nat_T1} is estimated by the value proportional to $\bar \alpha$, and therefore constant $c_1$ is estimated by the value proportional to $\tau_T$. This result corresponds to the result of Theorem~\ref{Nat_T11}.
\begin{Theorem}\label{Nat_T12}
Let the conditions {\bf A1}, {\bf A2} be satisfied, $0<\alpha_t \leq \bar \alpha$, in the averaged discrete system~\eqref{Nat_10_Zt} the $\frac{\varepsilon}{4}$-consensus is achieved for time $T$, and for constants $c_1,\;c_2$ from Theorem~\ref{Nat_T1} the following estimate holds
\begin{equation}
c_{1}\tau_{T} e^{c_2 \tau_{T}^2 }\bar \alpha \leq \frac{\varepsilon}{4},
\end{equation}
{\bf then} the mean square $\varepsilon$-consensus in the stochastic discrete system~\eqref{Nat_8} at time $t$ is achieved.
\end{Theorem}
%Proof is in the Appendix.

Consider the important case where $\forall i \in N \; f^i(x, u) = u$ and $\alpha_t=\alpha=const$. In this case the discrete averaged system~\eqref{Nat_10_Zt} has the form:
 \begin{equation}
\label{Nat_T8dicr}
 \bar Z_{t+1} = (I - ((I-U) -{\cal L}( \alpha A_{\max})) )Z_t.
 \end{equation}

\begin{Theorem} \label{Nat_T22new}
If conditions {\bf A2}, {\bf A3} are satisfied, $\alpha_t=\alpha>0$, $f^i(x,u)=u$ for any $i \in N$, and condition $\alpha < \frac{1}{d_{\max}}$ for matrix ${A_{\max}}$ is satisfied,
{\bf then}
the asymptotic mean square $\varepsilon$-consensus in the averaged discrete system~\eqref{Nat_T8dicr} is achieved. %JYM
%the asymptotic mean square $\varepsilon$-consensus in averaged discrete system~\eqref{Nat_T8dicr} for $n$ agents is achieved.

In addition, if the $\frac{\varepsilon}{4}$-consensus is achieved for the time $T(\frac{\varepsilon}{4})$  in the averaged discrete system~\eqref{Nat_T8dicr} and there exists $T_0>T(\frac{\varepsilon}{4})$  for which the parameter $\alpha$ provides the condition
\begin{equation}
\bar C_{1} e^{\bar C_2 } \alpha \leq \frac{\varepsilon}{4},
\end{equation}
$$
\bar C_1 = 8 n  \left(\tilde c  + \hat c(\frac{  \alpha^2 \tilde c}{c_3}+||\bar X_0||^2) e^{T\ln (c_3+1)}\right)\tau_{t},
$$
$$
\bar C_2 =  2^{2-\bar d}  \alpha^2||{\cal L}(A_{\max})||_2^2,\;\tilde c = n^2  \bar b^2 \sigma_w^2,\;\hat c = 2  n (n-1) \bar b^2 \tau_{t}^2 ,
$$
$$
c_3 =  2^{1+\tilde d}+2{\alpha^2}(||{\cal L}(A_{\max})||_2^2 +\hat c),
$$
where
$\tilde d = 0$ if $\bar d =0$ or $\tilde d = 1$ if $\bar d >0$,
 {\bf then} the mean square $\varepsilon$-consensus at time $t: \; T(\frac{\varepsilon}{4})\leq t \leq T$ in the stochastic discrete system~\eqref{Nat_8} is achieved.
\end{Theorem}
Note that in \cite{Huang}, under certain assumptions similar to the conditions of Theorem~\ref{Nat_T22new}, the necessary and sufficient condition for achieving the mean square consensus in case when the step sizes $\alpha_t$ tend to zero and the second term of (\ref{Nat_1}) has a simple form: $f^i(x_t^i, u_t^i)=u_t^i$ were proved.
%Выше был рассмотрен более общий случай вида функций $f^i(x_t^i, u_t^i)$ и не стремящиеся к нулю $\alpha_t$.
However, in the analysis above, the more general case of the form of functions $f^i(x_t^i, u_t^i)$ and step sizes $\alpha_t$ nondecreasing to zero has been considered. %was considered above. %JYM

%In our case we can also assume that the step size tends to zero. If  standard stochastic approximation  step size conditions are satisfied: $\sum_{t=0}^{\infty} \alpha_t = \infty$ and $\sum_{t=0}^{\infty} \alpha_t^2 < \infty$ then the time horizon of the continuous system is not bounded since $\tau_{t} =\alpha _{0} +\alpha _{1} +\ldots +\alpha _{t-1}$. %JYM

%%%%%%%%%%%%%%%%%%%%%%%%%%%%%%%%%%%%%%%%%%%%%%%%%%%%%%%%%%%%%%%%%%%%%%%%%%%%%%%%%%%%%%%%%%%%%%%%%%%%%%%%%%
%\section{Sensor Networks}
\section{The Load Balancing Problem}

To demonstrate the use of the results derived in the previous section, the load balancing problem is considered in this section.

%%%%%%%%%%%%%%%%%%%%%%%%%%%%%%%%%%%%%%%%%%%%%%%%%%%%%%%%%%%%%%%%%%%%%%%%%%%%%%%%%%%%%%%%%%%%%%%%%%%%%%%%%%
\subsection{Problem Statement}

In recent years, distributed parallel computing systems have been increasingly used \cite{AmelinNeurocomp12}. For such systems the problem of separating a package of jobs among several computing devices is important. Similar problems arise also in transport networks \cite{GranichinScobelev12, CAP12_2} and in production networks \cite{Armbruster}.

%We consider the sensor network of $n$ sensors that collect different type of information with feedback. Denote $N=\{1,\ldots,n \}$ as a set of agents (sensors, nodes), each of which can collect the information, get the information from its neighbors and send the information to server. The information can be redistributed among agents.
We consider a system that separates the same type of jobs among different agents, for parallel computing or production with feedback. Denote $N=\{1,\ldots,n \}$ as the set of intelligent agents, each of which serves the incoming requests using a first-in-first-out queue. Jobs may be received at different times and by different agents.

At any time $t$, the state of agent $i$, $i=1,\ldots,n$ is described by two characteristics:

\begin{itemize}
%\item $q_{t}^{i} $ is the information flow of sensor $i$ at time $t$;
%\item $p_{t}^{i} $ is the energy level of sensor $i$ at time $t$.
\item $q_{t}^{i} $  is the queue length of the atomic elementary jobs of the agent $i$ at time $t$;
\item $p_{t}^{i} $  is the productivity of the agent $i$ at time $t$.
\end{itemize}

The dynamics of each agent are described by
\begin{equation}
\label{Nat_11}
%p_{t+1}^{i}= p_{t}^{i} - r ln(t) ;\; \; i \in N,\; t = 0,  1, \ldots, T,
q_{t+1}^{i} =q_{t}^{i} -p_{t}^{i} + z_{t}^{i} + u_{t}^{i} ;\; \; i \in N,\; t = 0,  1, \ldots, T,
\end{equation}
%$$q_{t+1}^{i} =q_{t}^{i} + z_{t}^{i} + u_{t}^{i} ;\; \; i \in N,\; t = 0,  1, \ldots, T,$$
%where $r$ is the rate of battery discharge, $z_{t}^{i} $ is the new information received by agent $i$ at time $t$,  $u_{t}^{i}$ is the result of information redistribution between agents, which is obtained by using the selected protocol of information redistribution. In the dynamics we assume that $\sum_i u_{t}^{i} = 0, \; t = 0, 1, 2, \ldots$.
where $z_{t}^{i} $ is the new job received by agent $i$ at time $t$, $u_{t}^{i}$ is the result of jobs redistribution between agents, which is obtained by using the selected protocol of jobs redistribution. In the dynamics we assume that $\sum_i u_{t}^{i} = 0, \; t = 0, 1, 2, \ldots$.

We assume, that each agent $i \in N$ at time $t$ can receive the following information to form the control strategy:
\begin{itemize}
%\item the noisy observations about its information flow
\item noisy observations about its queue length
\begin{equation}
\label{Nat_3ii}
y_{t}^{i, i} = q_{t}^{i} + w_{t}^{i, i},
\end{equation}
%\item the noisy and delayed observations about its neighbors information flow, if $N_{t}^{i} \ne \emptyset$
\item noisy and delayed observations about its neighbors' queue length, if $N_{t}^{i} \ne \emptyset$
\begin{equation}
\label{Nat_3b}
y_{t}^{i, j} = q_{t-d_{t}^{i,j} }^{j} + w_{t}^{i, j},\; j \in N_{t}^{i},
\end{equation}
where $w_{t}^{i,j} $ are noises, $0 \le d_{t}^{i,j} \le \bar d$ is an integer-valued delay, and $\bar d$ is a maximal delay,
%\item the information about its energy level $p_{t}^{i}$ and about its neighbors energy level $p_{t}^{j},~~j \in N_t^i$.
\item information about its productivity $p_{t}^{i}$ and about its neighbors' productivities $p_{t}^{j},~~j \in N_t^i$.
\end{itemize}

%JYM
Let the fraction $\frac{q_{t}^{i}}{p_{t}^{i}}$ denote \textit{the load} of agent $i$ at time $t$, and $T_t$ denote the implementation time of jobs at time $t$, where
\begin{equation}
 T_{t} =\mathop{\max}_{i \in N} \frac{q_{t}^{i}}{p_{t}^{i}}.
\end{equation}
The objective is to balance the load such that the implementation time can be minimized.

\subsection{Control and Analysis}

To achieve the goal it is natural to use a redistribution protocol for jobs over time. Let's consider a special case where all jobs come to different agents at the initial time and no new job is received later. For this case, we have the following results.

\begin{Lemma}
\label{lemma_optim}
\textit{\textbf{(about the optimal control strategy)}}\
For the special case, among all possible options for redistributing jobs, the minimum completion time is achieved when
%наименьшее время работы системы соответствует тому, при котором
\begin{equation}q_t^i/p_t^i = q_t^j/p_t^j, \; \forall i, j \;\in N.\end{equation}
\end{Lemma}

\begin{Corollary} %JYM Collorary
If we take $x_t^i=q_t^i/p_t^i$ as the state of agent $i$  in a dynamic network, then the control goal --- to achieve consensus in the network, will correspond to the optimal job redistribution between agents in the special case.
\end{Corollary}

These above results imply that the load balancing problem can essentially be treated as a consensus problem, i.e. how to keep the load equal among all agents in the network. We highlight that for this special case the form~(\ref{Nat_11}) corresponds to the difference equation~(\ref{Nat_1}).

Based on this intuition, we extend to the more general case where new jobs may arrive to any of the $n$ agents at any time $t$. Specifically, consider the control protocol~(\ref{Nat_7}), where $\forall\; i \in N$, $\forall\; t $
denote $\bar N_t^i = N_t^i$ and $b_{t}^{i,j} = p_t^j / p_t^i, \; j \in N_t^i$. Here, we assume that $p_t^i \neq 0 \forall \; i$. Then, the dynamics of the load-balancing system~\eqref{Nat_11} with local voting protocol~(\ref{Nat_7}) is as follows:
\begin{equation}
\label{Nat_12}
%x_{t+1}^{i} =x_{t}^{i} + z_{t}^{i}/p_{t}^{i} +\alpha _{t} \sum _{j\in N_{t}^{i} } b_{t}^{i,j} (y_{t}^{i,j}/ p_t^j - y_{t}^{i,i}/ p_t^i ).
x_{t+1}^{i} =x_{t}^{i} - 1 + z_{t}^{i}/p_{t}^{i} +\alpha _{t} \sum _{j\in N_{t}^{i} } b_{t}^{i,j} (y_{t}^{i,j}/ p_t^j - y_{t}^{i,i}/ p_t^i ).
\end{equation}
where $\alpha _{t}$ are step sizes of control protocol, $y_{t}^{i,j} $ are noisy and delayed observation about $j$-th agents queue length, $z_{t}^{i} $ is the new job received by agent $i$ at time $t$.

Consequently, results of consensus achievement in the previous section apply. Particularly, if the graph is balanced, for the general setting with random uncertainties in the measurements, in the network topology, and in the protocol control (\ref{Nat_7}), Theorem~\ref{Nat_T22new} allows to reduce the study of the dynamics of the load balancing system to the investigation of the corresponding averaged discrete model.

\begin{Theorem}\label{Nat_T3}
%Если данные о производительности узлов с течением времени стабилизируются: $\exists \;\lim_{t \to \infty} E r_t^i = \bar r^i >0$, $\forall \; i \in N$, и выполнены условия {\bf A0}, {\bf A2а--г},
%{\bf тогда}
%$n$ узлов достигают асимптотического среднеквадратичного $\varepsilon$-консенсуса с $\varepsilon=C_3{\bar \alpha}^{\mu }$ и константами $C_3, {\bar \alpha}, {\mu }$ из формулировки теоремы~\ref{Nat_T1}.
If $\alpha_t = \alpha = const$ is sufficiently small, the productivities stabilize over time: $\exists \; \mathrm{E} p_t^i = \bar p^i >0$, $\forall \; i \in N$, conditions {\bf A2}, {\bf A3} and condition~\eqref{Nat_dmax} for matrix ${A_{\max}}$ are satisfied, in the averaged discrete system the $\frac{\varepsilon}{4}$-consensus is achieved for the time~$T(\frac{\varepsilon}{4})$,
%, $T<T(\frac{\varepsilon}{4})$
and there exists $T_0>T(\frac{\varepsilon}{4})$ for which the parameter $\alpha$ ensures the condition
\begin{equation}
\bar C_{1} e^{\bar C_2 } \alpha \leq \frac{\varepsilon}{4},
\end{equation}
where
$$
\bar C_1 = 8 n  \left(\tilde c  + \hat c(\frac{  \alpha^2 \tilde c}{c_3}+||\bar X_0||^2) e^{T\ln (c_3+1)}\right)\tau_{t},
$$
$$
\bar C_2 =  2^{2-\bar d}  \alpha^2||{\cal L}(A_{\max})||_2^2,\;\tilde c = n {(\sigma_w/\bar p^i)}^2 \bar b ,\;\hat c = 2  n \bar b \tau_{t} ,
$$
$$
c_3 =  2^{1+\tilde d}+2{\alpha^2}(||{\cal L}(A_{\max})||_2^2 +\hat c),
$$
$\tilde d = 0$ if $\bar d =0$, or $\tilde d = 1$ if $\bar d >0$,
{\bf then}
in the stochastic discrete system for the $n$ agents at time $t: \; T(\frac{\varepsilon}{4})\leq t \leq T$, the $\varepsilon$-consensus is achieved.
\end{Theorem}

We remark, that in Theorem~\ref{Nat_T3}, the conditions for productivities of agents are rather general. They hold for an adaptive problem statement, when information about the actual productivities is specified over time. In addition, due to the fact that the step sizes $\alpha_t$ of the control protocol~(\ref{Nat_7}) do not tend to zero, the considered control protocol shows good performance in the more general problem case. In a number of similar cases the validity of applying stochastic approximation control strategies  with non-decreasing to zero step sizes in nonstationary problems could be theoretically proved (see, e.g., \cite{Granichin, Vakhitov, Granichin2}).

\subsection{Simulation Results}

\subsubsection{The six-node case}

To show the convergence to consensus and to compare the initial stochastic system with the averaged model, we give an example of simulation for a computer network consisting of six computing agents. %JYM

\begin{figure}[thpb]
      \centering
      \includegraphics[scale=0.2]{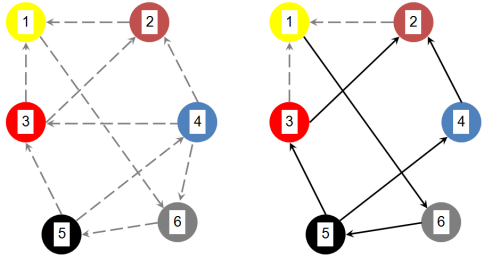}
      \caption{Maximal set of communication links $E_{\max}$ (left Fig.); Network topology at time $t$ (right Fig.).}
      \label{e}
\end{figure}

Fig. \ref{e} (left) shows the network, indicating the possible communication links, some of which may be ``closed'' and ``opened up'' over time. The network topology is random at any time $t$, and particularly, Link 1-3 or 1-2 appears with probability $1/2$ (Fig. \ref{e} (right)). %JYM

For the case without delay, equation~\eqref{e2} is as follows:
 \begin{equation}
\label{Nat_10-2}
\frac{ d X}{d \tau}=R(\alpha, \bar x),
\end{equation}
where
\begin{equation}
\label{Nat_14}
R(\alpha, \bar x)=\left(\begin{array}{cccccc}
{-1} & {\frac{1}{2} \frac{p^{2}}{p^{1}} } & {\frac{1}{2} \frac{p^{3}}{p^{1}}} & {0} & {0} & {0}
\\ {0} & {-1} & {0} & {\frac{p^{4}}{p^{2}}} & {0} & {0}
\\ {0} & {0} & {-1} & {0} & {\frac{p^{5}}{p^{3}}} & {0}
\\ {0} & {0} & {0} & {-1} & {\frac{p^{5}}{p^{4}}} & {0}
\\ {0} & {0} & {0} & {0} & {-1} & {\frac{p^{6}}{p^{5}}}
\\ {\frac{p^{1}}{p^{6}}} & {0} & {0} & {0} & {0} & {-1}
\end{array}\right).
\end{equation}

In the case of uniformly distributed delays in the measurements, where the integer-valued delay $d_{t}^{ij} $ equals 0 or 1 with probability 1/2, $\bar d =1, p_0^{ij}=p_1^{ij}=1/2$, we extend the state space:
\begin{equation}
\bar X_{t} =[x_{t}^{1} ,\ldots ,x_{t}^{n} ,x_{t-1}^{1} ,\ldots x_{t-1}^{n} ]\in {\mathbb R}^{2n}. \end{equation}

Matrix $G$ of the corresponding averaged discrete model~\eqref{Nat_10_Zt} is as follows:
\begin{equation}
 G=\left(\begin{array}{cc} {\frac{1}{2} H \alpha} & {\frac{1}{2} H \alpha} \\ 0 & 0 \end{array}\right),
\end{equation}
where
\begin{eqnarray}
H=\left(\begin{array}{cccccc}
{0} & {\frac{1}{2} \frac{p^{2}}{p^{1}}} & {\frac{1}{2} \frac{p^{3}}{p^{1}}} & {0} & {0} & {0}
\\ {0} & {0} & {0} & {\frac{p^{4}}{p^{2}}} & {0} & {0}
\\ {0} & {0} & {0} & {0} & {\frac{p^{5}}{p^{3}}} & {0}
\\ {0} & {0} & {0} & {0} & {\frac{p^{5}}{p^{4}}} & {0}
\\ {0} & {0} & {0} & {0} & {0} & {\frac{p^{6}}{p^{5}}}
\\ {\frac{p^{1}}{p^{6}}} & {0} & {0} & {0} & {0} & {0}
\end{array}\right).
\end{eqnarray}

%We carry out simulation for the system shown in Fig. \ref{e}. Take following initial node loads: $x_0^1=5000,~x_0^2=3500,~x_0^3=2300,~x_0^4=3150,~x_0^5=7400,~x_0^6=1100$. Productivity of the nodes: $p^1=2,~p^2=0.75,~p^3=1.2,~p^4=1.7,~p^5=3.5,~p^6=2.1$ and they are not changing in time.

We set the initial queue lengths and the productivities of agents, and assume that the productivities of nodes do not change over time. In addition, we highlight that the information about the queue lengths is measured with random noise and delays.
% as: $q_{0}^{1} =5000,\; q_{0}^{2} =3500,\; q_{0}^{3} =2300,\; q_{0}^{4} =3150,\; q_{0}^{5} =7400,\; q_{0}^{6} =1100$ and the productivities of agents: $p^{1} =2,\; p^{2} =0.75,\; p^{3} =1.2,\; p^{4} =1.7,\; p^{5} =3.5,\; p^{6} =2.1$. We assume that the productivities of nodes do not change over time.

\begin{figure}[thpb]
      \centering
      \includegraphics[width=0.32\textwidth]{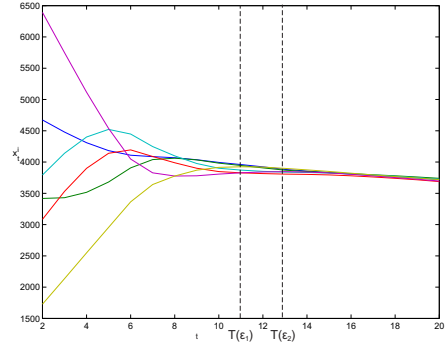}
      \caption{Dynamics of the agents $x_{t}^{i}$ at the start and time to consensus}
      \label{time}
\end{figure}

\begin{figure}[thpb]
      \centering
\subfigure[Dynamics of the agents $x_{t}^{i}$]{\includegraphics[width=0.24\textwidth]{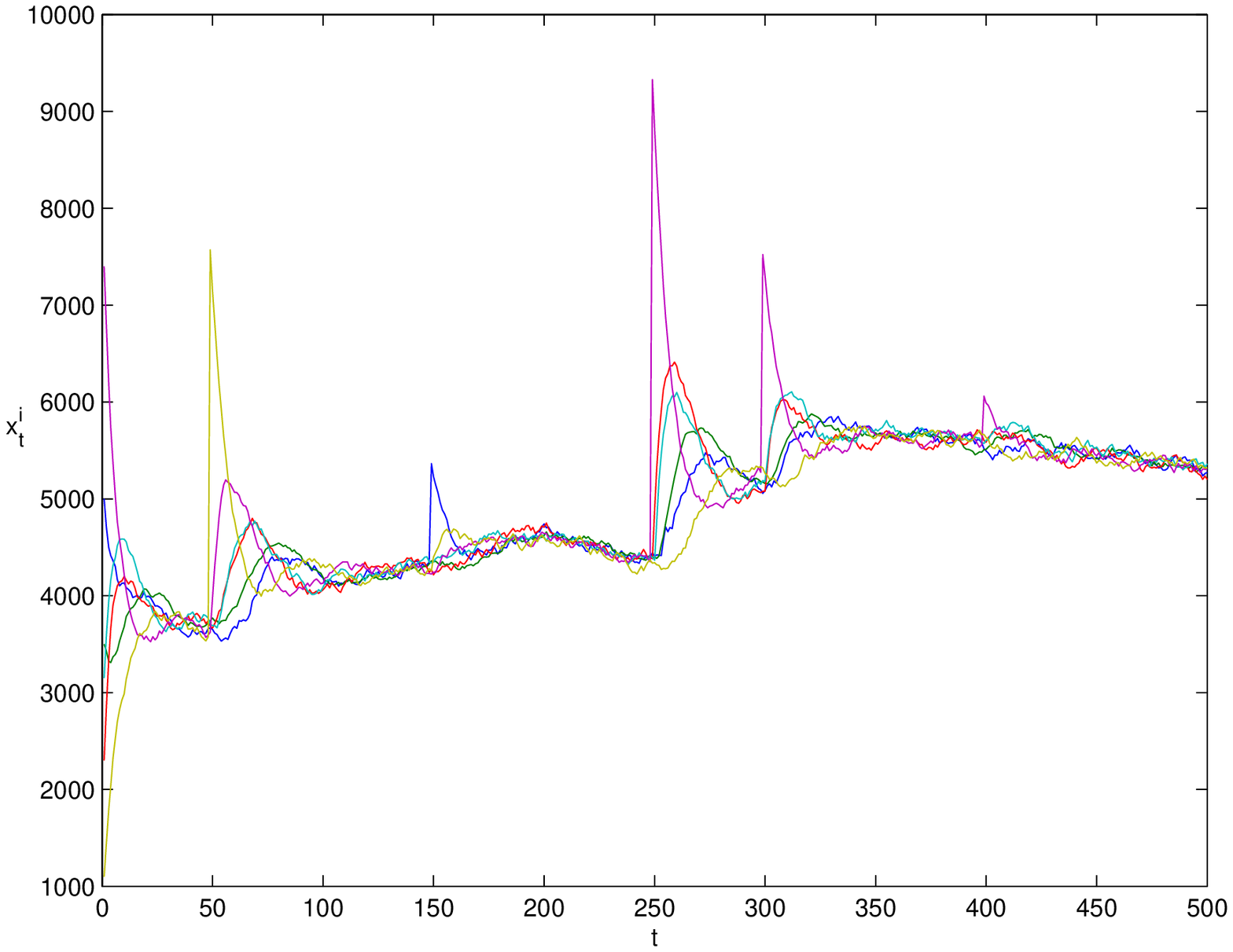}}
\subfigure[Convergence to consensus]{\includegraphics[width=0.24\textwidth]{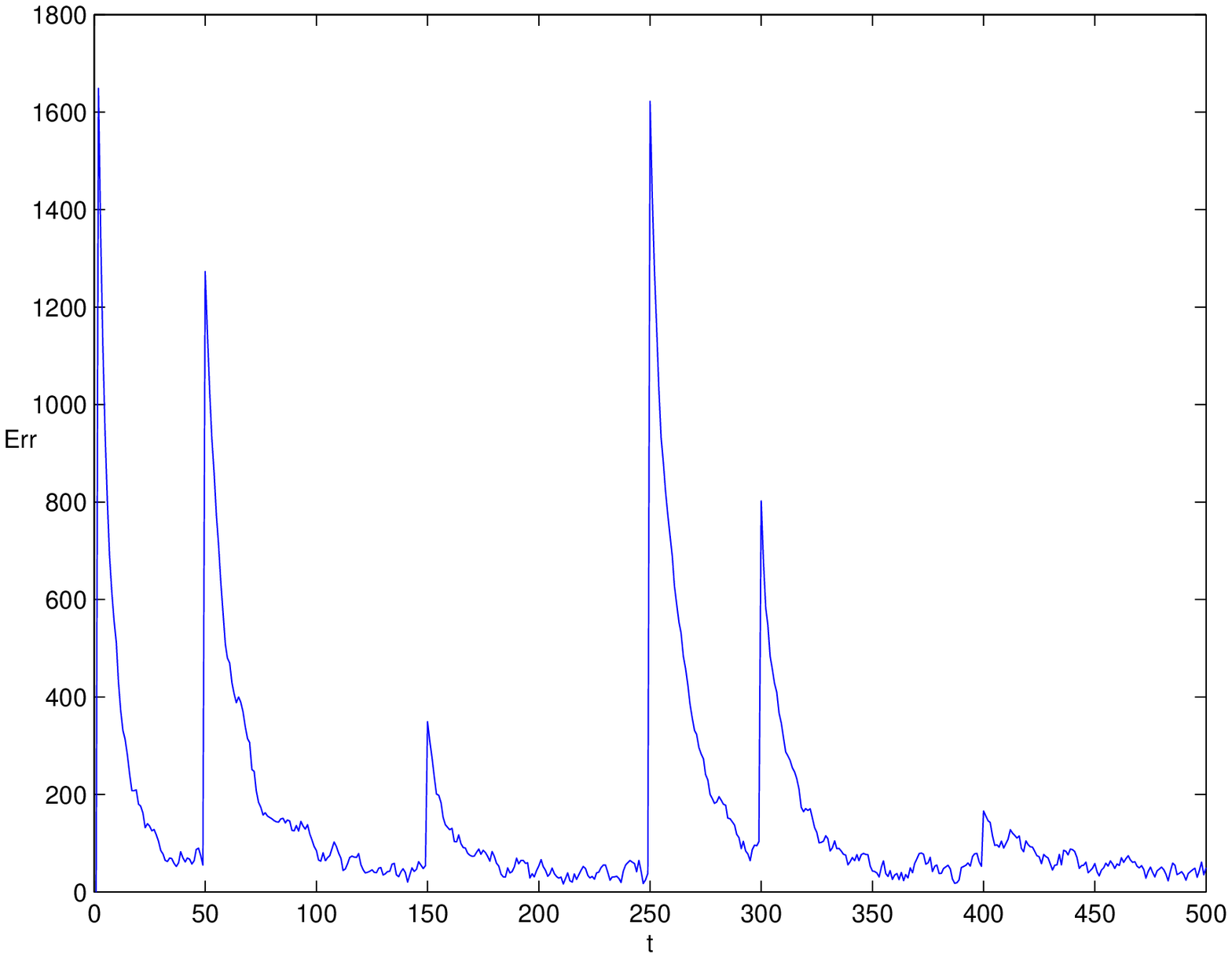}}
      \caption{The 6-node general case}
      \label{nonsatat_case}
\end{figure}

We consider two cases, the special case and the general case, as discussed in the previous subsection. We use constant step size $\alpha_{t}=\alpha =0.1$. The dynamics of the agents $x_{t}^{i} $ with local voting protocol~\eqref{Nat_12} is shown in Fig.~\ref{time} and~\ref{nonsatat_case}.

Fig.~\ref{time} shows how the system operates in the special case when there are no new incoming jobs during the system work (only the initial load). Each line, corresponding to one node, indicates how the load $x_t^{i}$ evolves over time. These lines also show how the system evolve to reach load-balancing or consensus.

Now we estimate the time to consensus. We calculate eigenvalues and
%$$
%eig = \large\left(\begin{array}{c}   0.0000 \\  -0.7737 + 0.9522i \\  -0.7737 - 0.9522i \\  -1.7281 + 0.3929i \\  -1.7281 - 0.3929i \\ -10.5000 \\ -10.1419 + 0.4882i \\ -10.1419 - 0.4882i \\  -9.6064 + 0.2235i \\  -9.6064 - 0.2235i \\  -1.0000 \\ -10.0000\end{array}\right)
%$$
obtain that $|Re(\lambda_2)|=0.7737$.
By formula~(\ref{NL_Teps}) we can calculate ${ T}(\varepsilon)$ for continuous system. If $\varepsilon = 0.1$ then ${ T}(\varepsilon) = 12.8883$. If $\varepsilon = 1$ then ${ T}(\varepsilon) = 11.4003$. The corresponding values are marked on Fig.~\ref{time}.

%Fig.~\ref{satat_case}.(a) is the start part of Fig.~\ref{satat_case}.(b), where each line, corresponding to one node,
%(with the same color in Fig. \ref{e}),
%Fig.~\ref{satat_case}.(a) shows the dynamics of the agents $x_{t}^{i}$ over time, where each line, corresponding to one node, indicates how the load $x_t^{i}$ evolves over time. These lines also show how the system evolve to reach load-balancing or consensus. Fig.~\ref{satat_case}.(b) shows, how the average residual changes over time: it rapidly reduces and as the time goes beyond 40, the residual retains at low level.

%\begin{figure}[thpb]
%      \centering
%      \includegraphics[scale=0.4]{dinshortsatatTE2.eps}
%%      \includegraphics[scale=0.6]{dinsatatE2.eps}
%      \includegraphics[scale=0.25]{dinsatat.eps}
%      \includegraphics[scale=0.25]{errstat.eps}
%      \caption{The dynamics of the agents $x_{t}^{i} $ and the average residual for stationary case}
%      \label{satat_case}
%\end{figure}

%\begin{figure*}[tb!]
%\centering
%\subfigure[Dynamics of the agents $x_{t}^{i}$ at the start and time to consensus]{\includegraphics[width=0.32\textwidth]{dinshortsatatTE2.eps}}
%\subfigure[Dynamics of the agents $x_{t}^{i}$ over time]{\includegraphics[width=0.32\textwidth]{dinsatat.eps}}
%\subfigure[Average residual]{\includegraphics[width=0.32\textwidth]{errstat.eps}}
%      \caption{The 6-node special case}
%      \caption{The dynamics of the agents $x_{t}^{i} $ in (a) and (b), and the average residual in (c)} % stationary case}
%      \label{satat_case}
%\end{figure*}

\begin{figure*}[thpb]
      \centering
\subfigure[All nodes]{\includegraphics[width=0.32\textwidth]{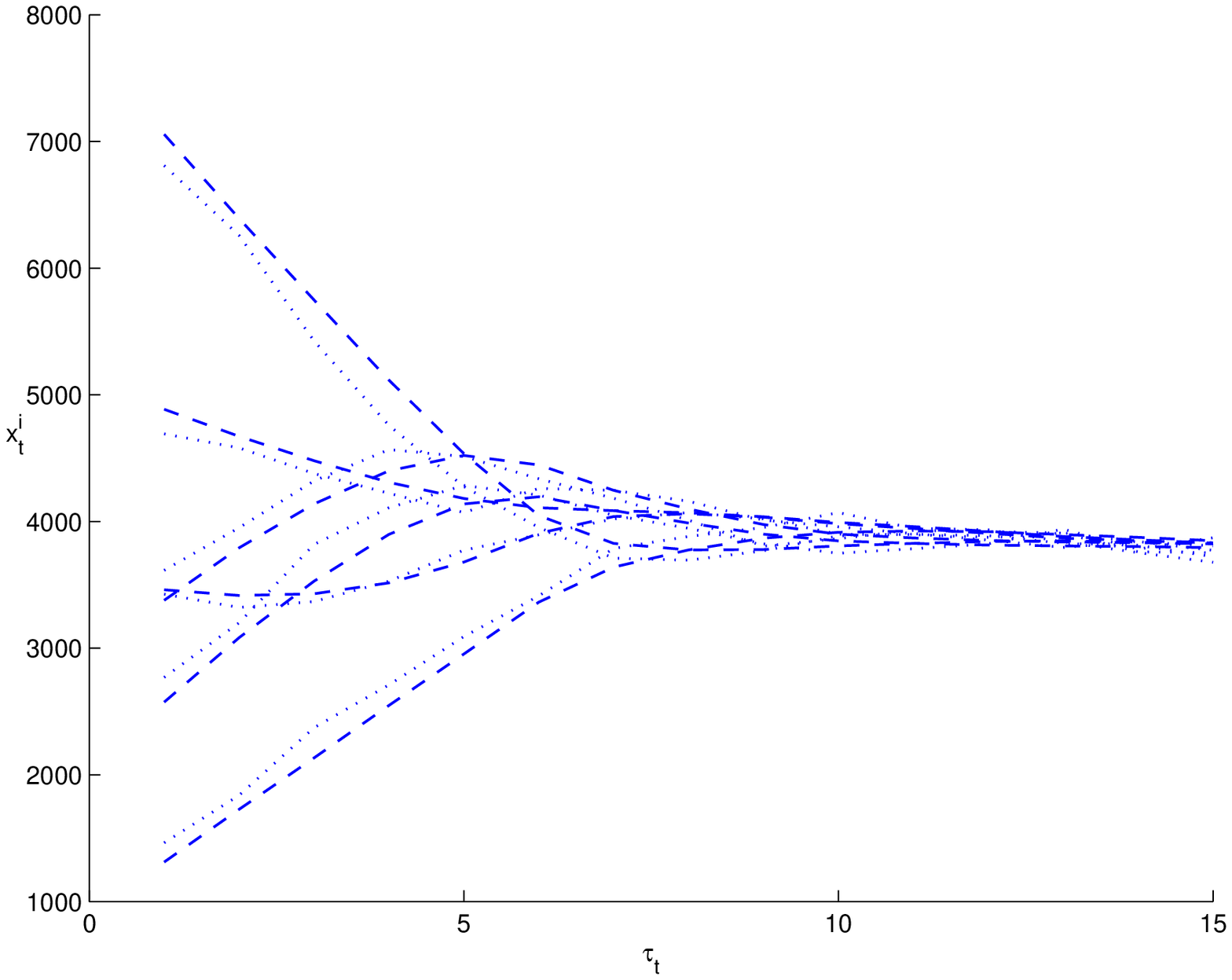}}
\subfigure[For the first node]{\includegraphics[width=0.32\textwidth]{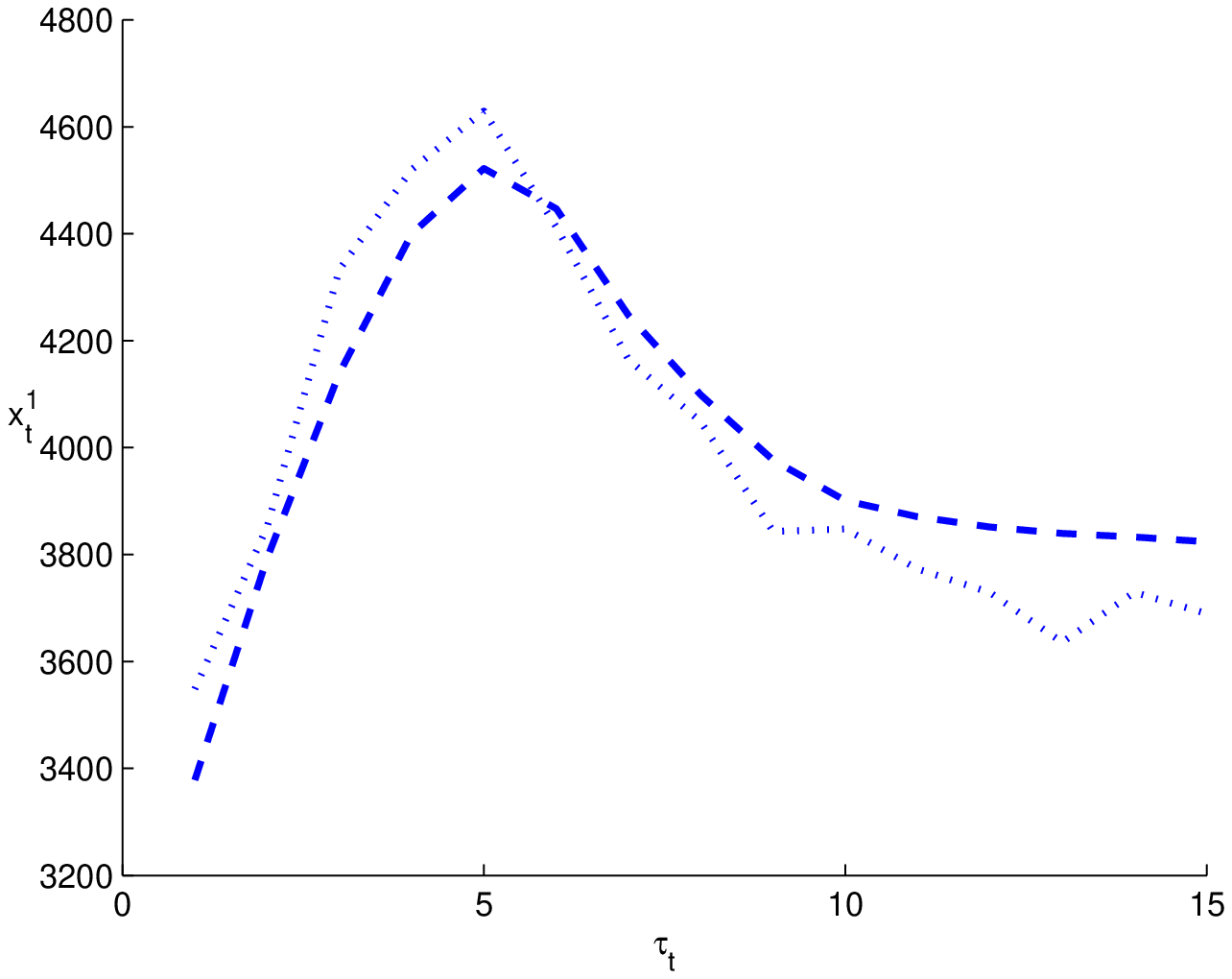}}
\subfigure[For the second node]{\includegraphics[width=0.32\textwidth]{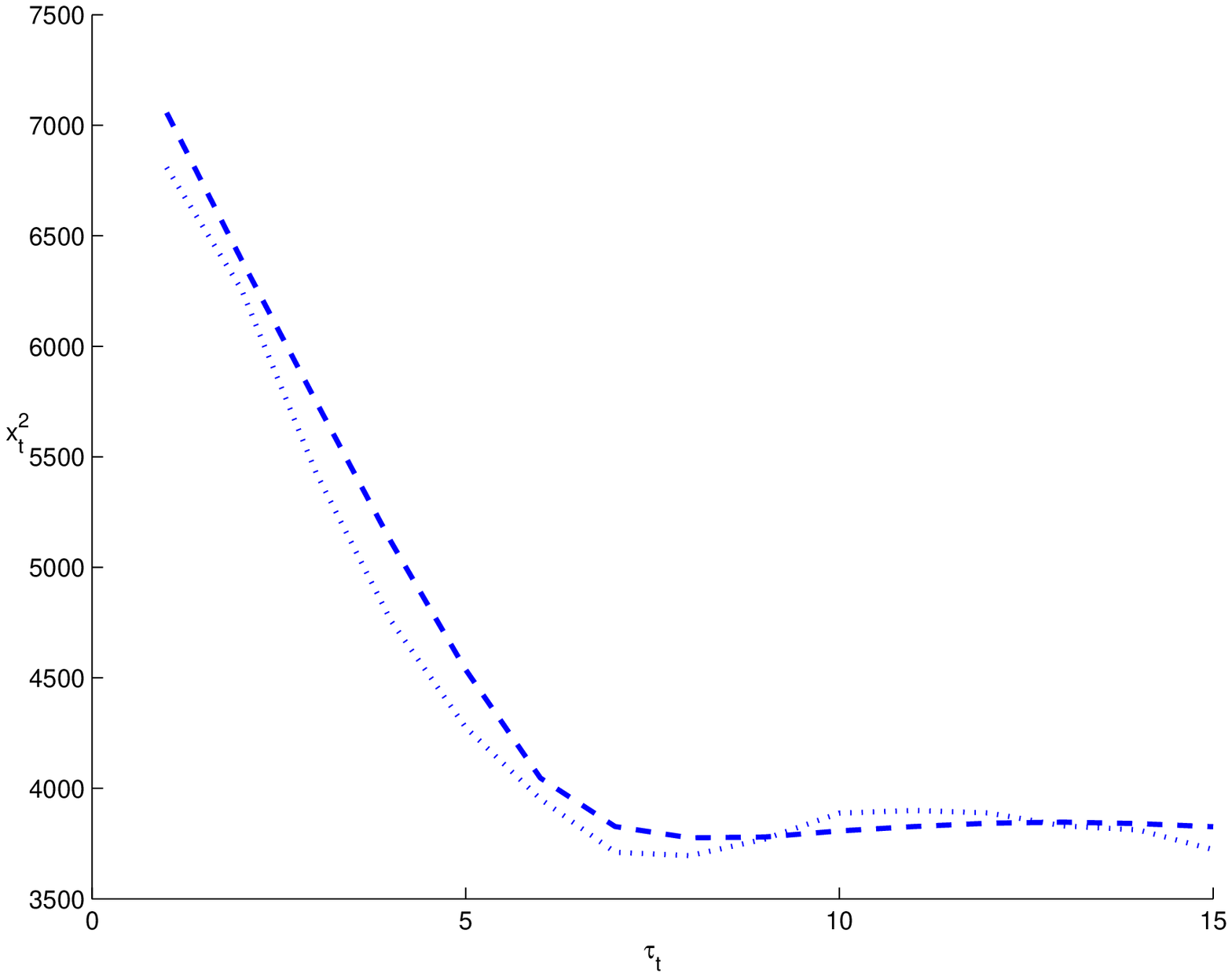}}
      \caption{Comparison of trajectories of the stochastic discrete system and its averaged model}
      \label{comp}
\end{figure*}

To support that we can use the averaged model to study our initial stochastic system, Fig. \ref{comp} is presented. The figure compares the dynamics of algorithm \eqref{Nat_7} and that of the averaged model described in Sec. \ref{sec-main}. Fig. \ref{comp} shows that trajectories of the stochastic discrete system (dotted lines) are close with the limiting trajectories of the average system (dashed lines).

To characterize the quality of the protocol~(\ref{Nat_7}) in terms of convergence of trajectories to consensus $x^{\star}$, we use the average residual, defined as $Err=\sqrt{\sum _{i} \frac{(x_{t}^{i} -x^{\star})^{2} }{n} } $.

%In Fig.~\ref{nonsatat_case} we can see the system operation in nonstationary case. It means that new jobs can come to different agents during the system work. We can see that the incoming jobs do not affect the quality of the system. It is a big advantage of the algorithm.
Fig.~\ref{nonsatat_case} shows the dynamics of the system in a more general 6-node case where new jobs can come to different agents during the system work. New jobs arrive at a random node at random times. Specifically, Fig.~\ref{nonsatat_case}.(a) indicates how the system tries to reach consensus using the local voting protocol~(\ref{Nat_7}) when there are new incoming jobs. In addition, the quality of the protocol~(\ref{Nat_7}) is indicated by Fig.~\ref{nonsatat_case}.(b), where the corresponding evolvement of average residuals is displayed. It shows, how the average residual changes over time: it rapidly reduces and retains at low level until new jobs received, and then it reduces again. The simulation results shows the good performance of the control protocol~(\ref{Nat_7}) in general case. This is explained by the properties of the stochastic approximation type algorithm with non-decreasing step, since each time instant when new jobs received might be considered as an initial time instant.
In a number of similar cases the validity of applying stochastic approximation control strategies  with non-decreasing to zero step sizes in nonstationary problems could be theoretically proved (see, e.g., \cite{Granichin, Vakhitov, Granichin2}).

%\begin{figure}[thpb]
%      \centering
%%      \includegraphics[scale=0.3]{satat_case.eps}
%      \includegraphics[scale=0.4]{dinnonsat.eps}
%      \caption{The dynamics of the agents $x_{t}^{i} $ for nonstationary case.}
%      \label{nonsatat_case}
%\end{figure}

%\begin{figure}[thpb]
%      \centering
%      \includegraphics[scale=0.4]{errnonsat.eps}
%      \caption{Convergence to consensus.}
%      \label{er}
%\end{figure}

%The quality of the protocol~(\ref{Nat_7}) (convergence of trajectories to consensus $x^{\star}$) is characterized by the average residual $Err=\sqrt{\sum _{i} \frac{(x_{t}^{i} -x^{\star})^{2} }{n} } $ (Fig.~\ref{satat_case} and Fig.~\ref{er}).

\begin{figure}[thpb]
      \centering
      \includegraphics[scale=0.16]{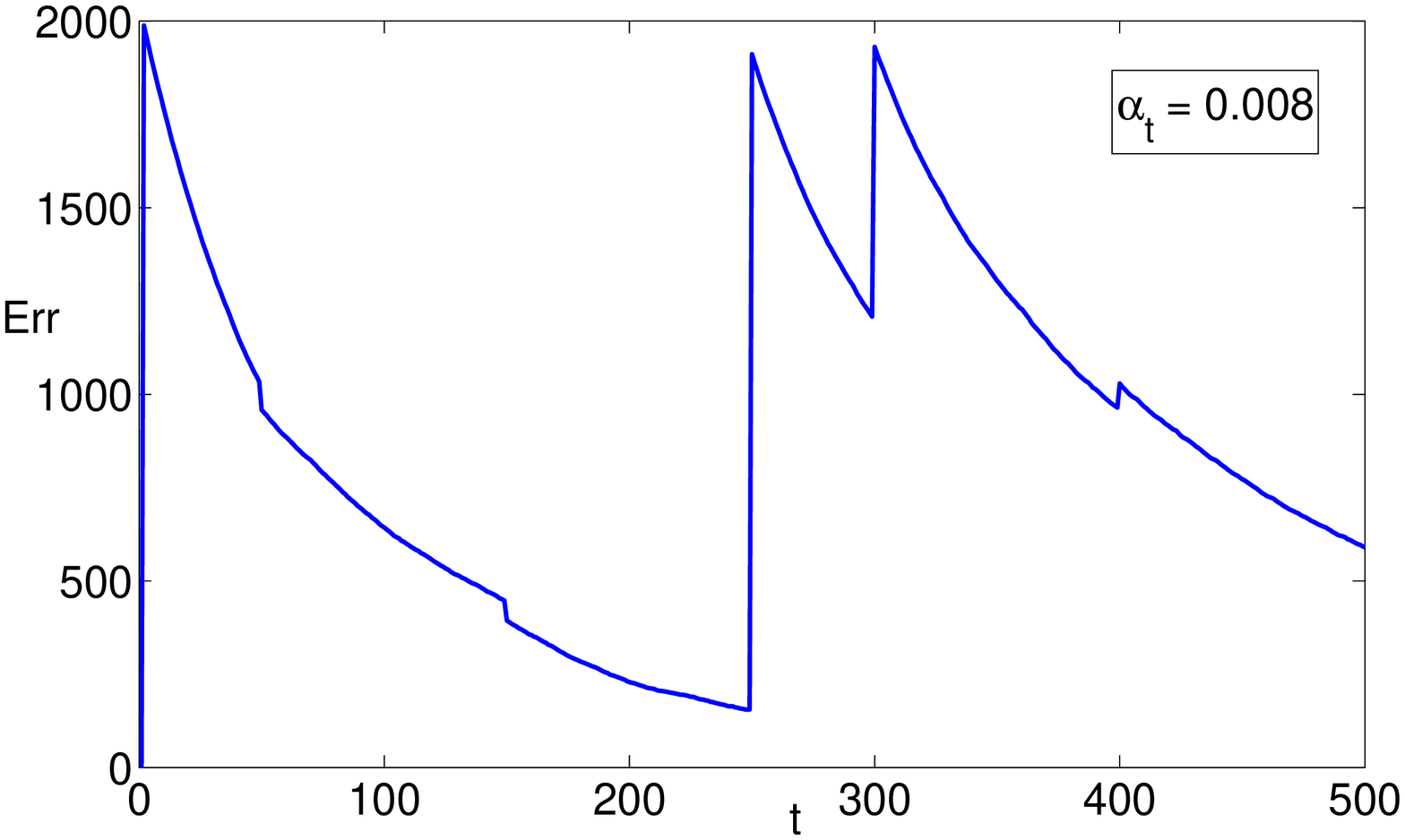}
      \includegraphics[scale=0.23]{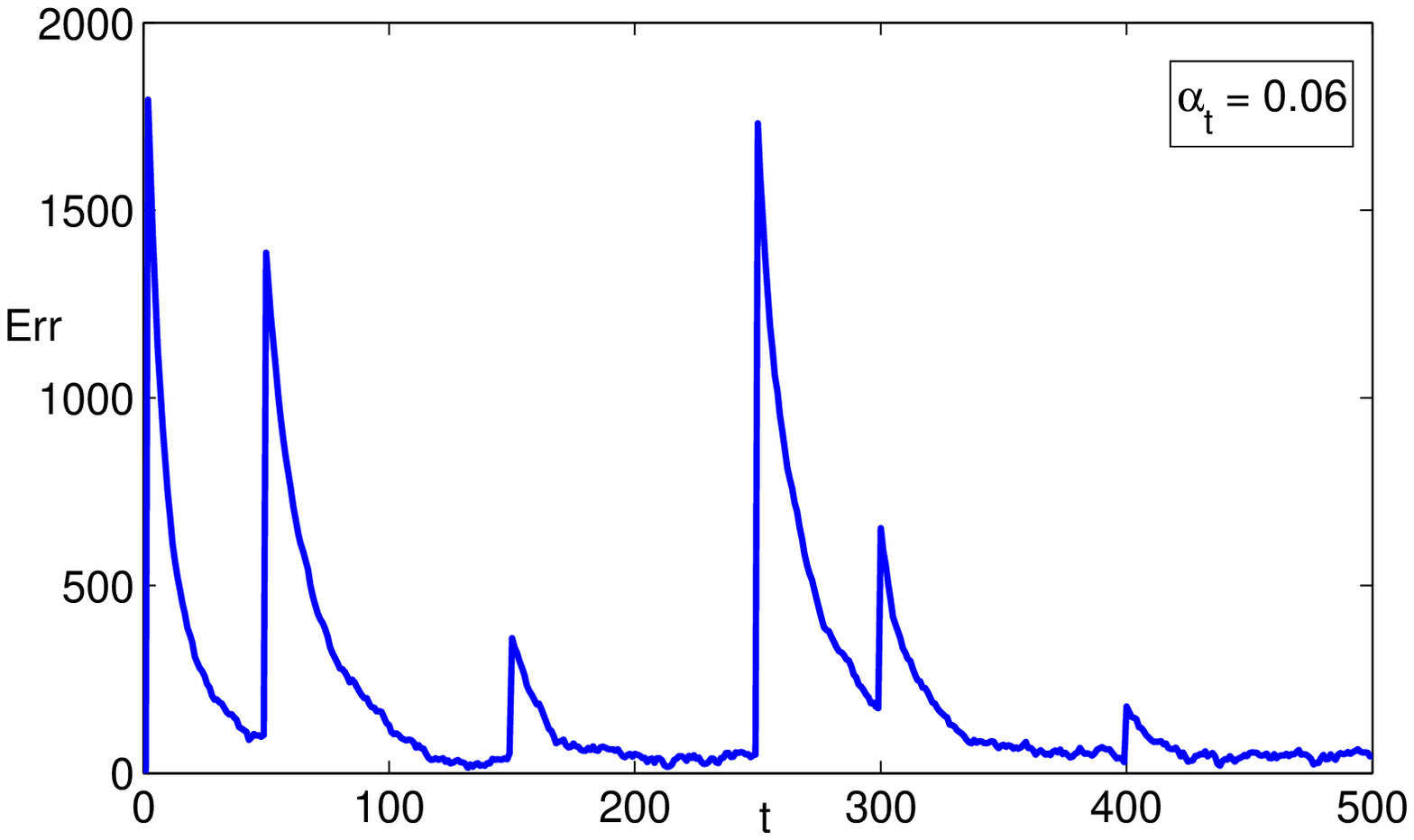}
      \includegraphics[scale=0.23]{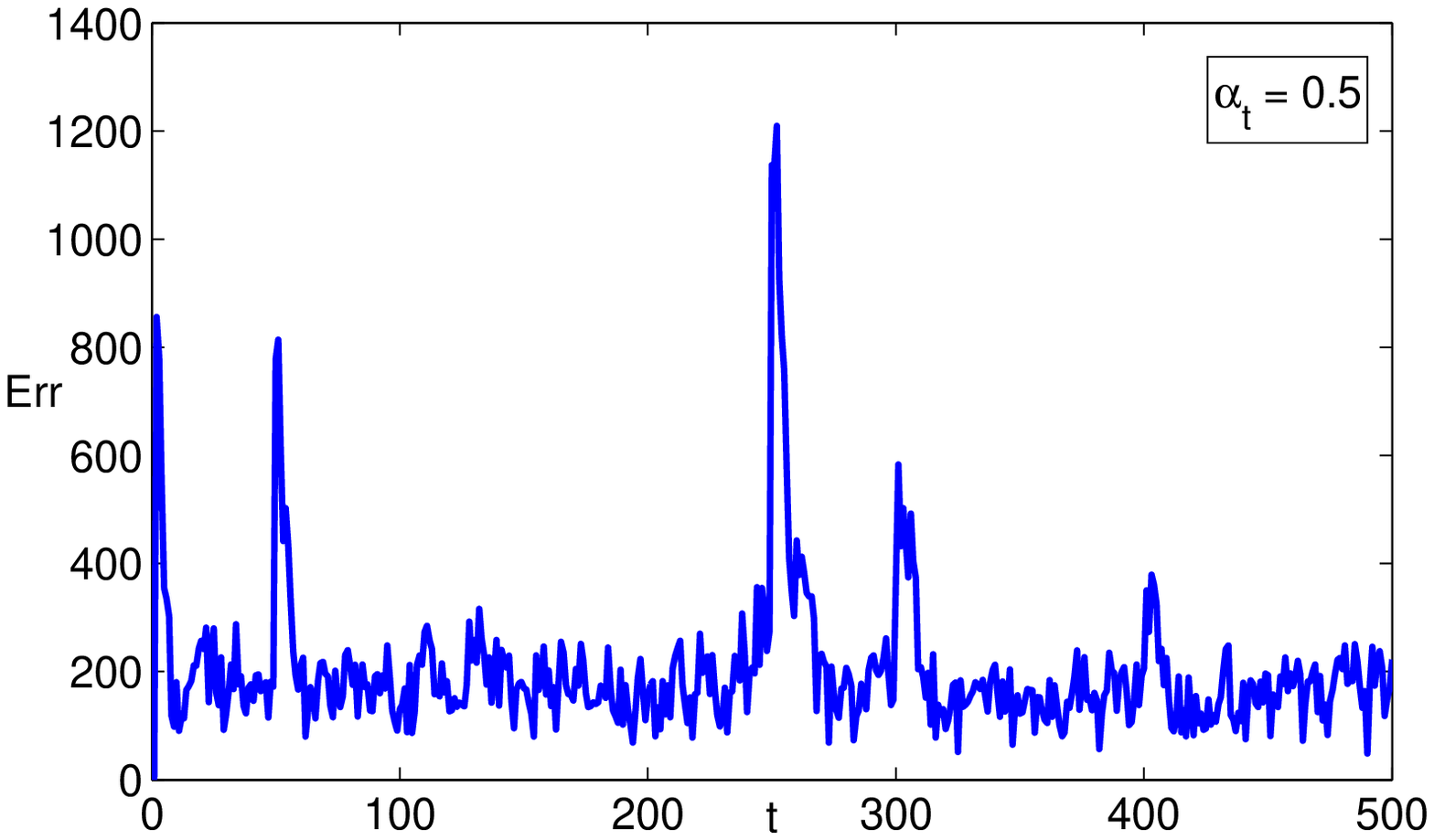}
      \includegraphics[scale=0.23]{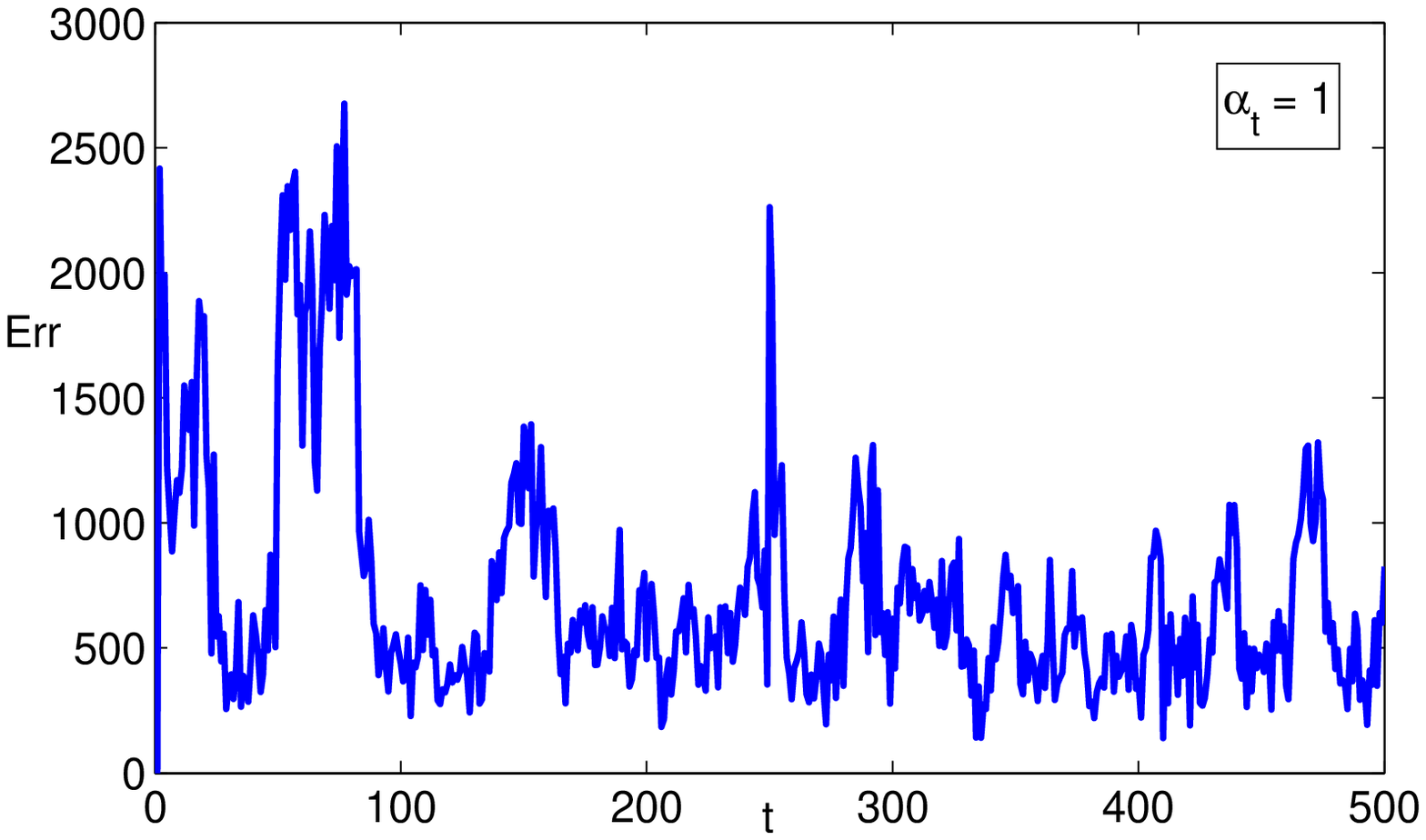}
      \includegraphics[scale=0.23]{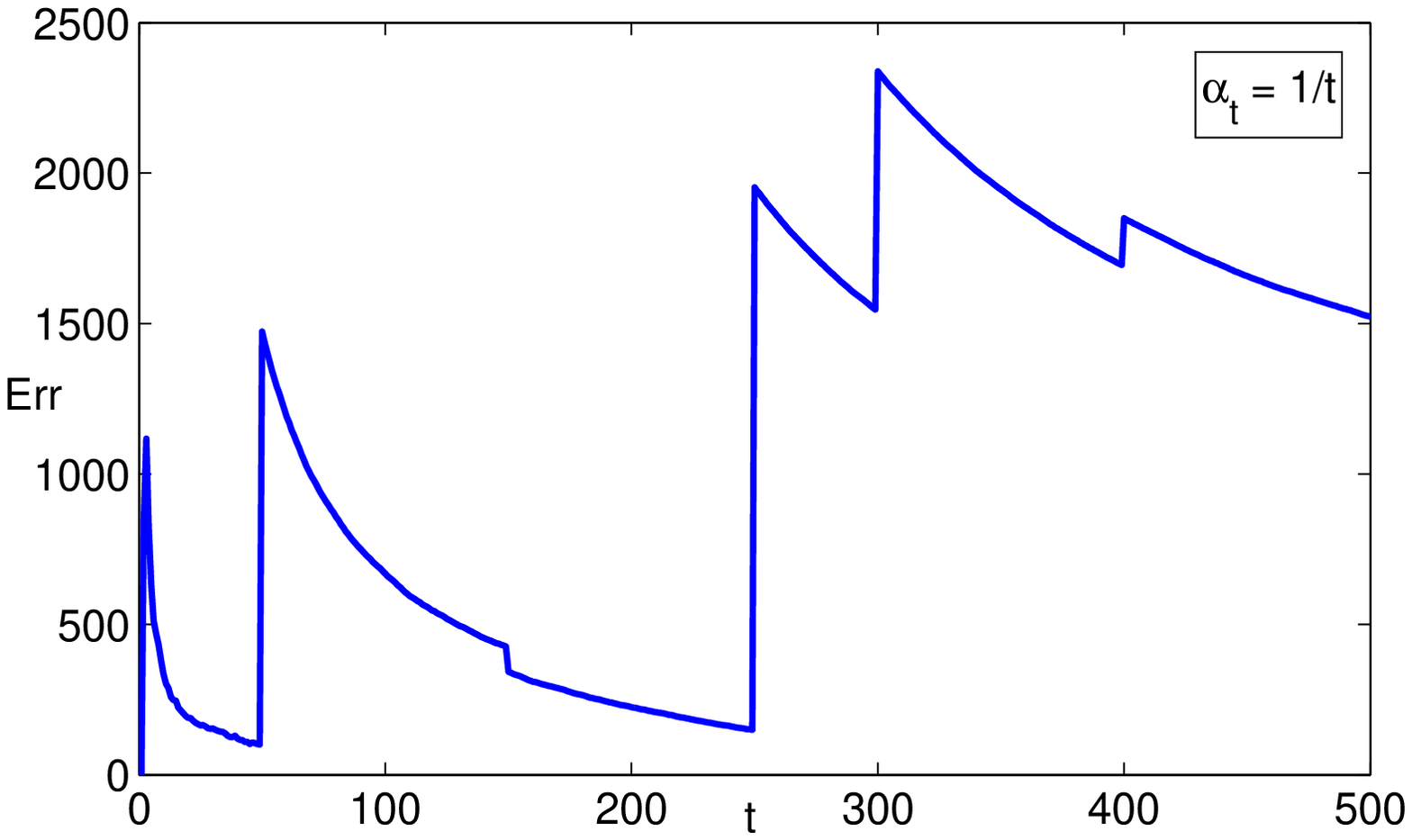}
      \caption{Convergence to consensus with different step sizes.}
      \label{alpha}
\end{figure}

In Fig.~\ref{alpha} there are graphs for the average residuals with using of different parameters of step sizes $\alpha$. In first four figures we used constant step sizes. It could be seen that if we increase the step size then the time to consensus will decrease until reaching a certain level. However, if we use the decreasing step size ($\alpha_t=1/t$) then the convergence rate  decreases with time.

%Compare the dynamics of algorithm \eqref{Nat_7} and of averaged model described above. %For simplicity, we assume that the system does not receive new orders and we have random delay.

%Fig. \ref{comp} shows that trajectories of stochastic discrete system (dotted lines) are close with the limiting trajectories of the average system (dashed lines).

%\begin{figure}[thpb]
%      \centering
%      \includegraphics[scale=0.4]{compxallstat.eps}
%      \includegraphics[scale=0.3]{compx1stat.eps}
%      \includegraphics[scale=0.24]{compx2stat.eps}
%      \caption{Trajectories of stochastic discrete system and its averaged model.}
%      \label{comp}
%\end{figure}

%The results from Section IV show that we can use the averaged model to study our initial stochastic system.

\subsubsection{The 1024-node case}

To show how well the approach works to achieve load balancing and the advantage of redistribution of jobs in a larger network, we consider a network of 1024 agents. The focus here is to compare the performance of the system adopting the local voting protocol~(\ref{Nat_7}) to redistribute load with that without load-redistribution.
%Agents are modeled as servers of the queuing system.

In the simulation, the time between events in the input stream is exponentially distributed with parameter $d_{in}= 1/3000$, and the normalized ``complexities'' of jobs are also exponentially distributed with parameter $d_{p}= 1$ (where, the normalized ``complexity'' of job is referred to as the time, required to perform the job on a single agent with productivity $p=1$). The number of incoming jobs is $10^6$. The choice of an agent, which receives the next job is performed randomly by the uniform distribution of 1024 agents.

\begin{figure}[hbp]
      \centering
      \includegraphics[scale=0.2]{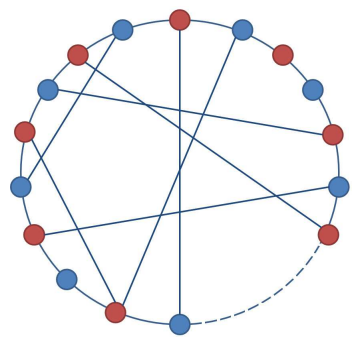}
      \caption{An example of network topology.}
      \label{graph}
\end{figure}

Agents are connected in a circle. In addition, there are $n$ random connections between agents on each iteration, that change over time. An example snapshot of the network is shown in Fig.~\ref{graph}.

Similarly, we also consider two cases. In the first, all jobs arrive at the initial time. In the second, the same jobs arrive at different time instants in the interval from 1 to 2000.

\begin{figure}[ht]
      \centering
      \includegraphics[scale=0.4]{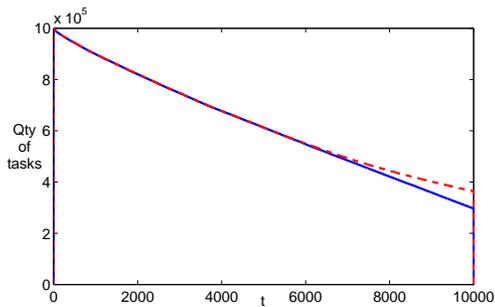}
      \caption{The number of jobs in the queue in case, when all jobs arrive at the initial time.}
      \label{Qty_tasks1}
\end{figure}

%In the first case all tasks arrive at different time instants in the interval from 1 to 2000.
For the first case, the randomization of nodes selection at the initial time provides a uniform load (load balancing) of all nodes in the beginning, but then the strategy without redistribution  begins to ``lose'' because the durations of jobs in the system are not known a priori, and some nodes start to ``slow down''. Fig.~\ref{Qty_tasks1} compares the number of jobs in queue with and without redistribution. Solid lines correspond to the case with redistribution of jobs by the local voting protocol, and dashed lines --- to case without redistribution. The figure also shows better performance achieved by using the local voting protocol.
%shows a graph of the number of tasks in the queue for the case when all tasks arrive at the initial time. Solid lines correspond to the case with redistribution of tasks by the local voting protocol, and dashed lines --- to case without redistribution.

%In the second case all tasks arrive at the initial time.

\begin{figure}[htb!]
      \centering
      \includegraphics[scale=0.4]{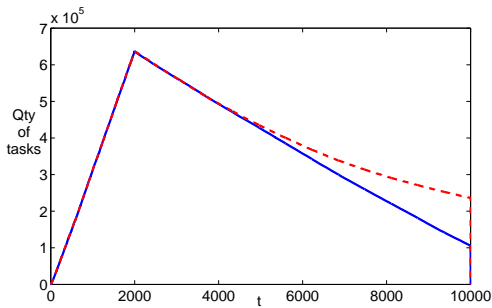}
      \caption{The number of jobs in the queue in case, when all jobs arrive at different time instants.}
      \label{Qty_tasks}
\end{figure}

\begin{figure}[htb!]
      \centering
      \includegraphics[scale=0.4]{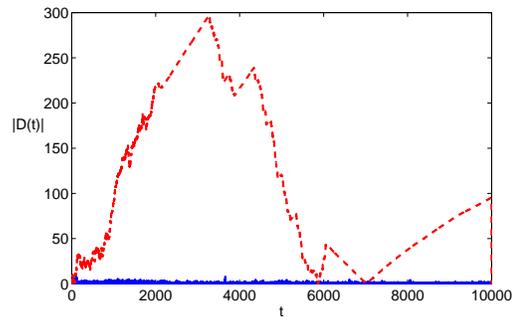}
      \caption{Maximum deviation from the average load on the network.}
      \label{mean_dev}
\end{figure}

In the second case, Figs.~\ref{Qty_tasks} and \ref{mean_dev} show typical results of simulations. %, when tasks were sent randomly to some agents at the feeded time.  %when they received.
In these figures, solid lines correspond to the case with redistribution of jobs by the local voting protocol, and dashed lines --- to the case without redistribution, where symbol $|D(t)|$ stands for the maximum deviation from the average load on the network. Figs.~\ref{Qty_tasks} and \ref{mean_dev} show that the performance of the adaptive multi-agent strategy with the redistribution of jobs among ``connected'' neighbors is significantly better than the performance of the strategy without redistribution.

\section{Conclusion}

In this paper, the approximate consensus problem statement of multi-agent stochastic system with nonlinear dynamics, noise, delays and switched topology was introduced. In contrast to the existing stochastic approximation-based control algorithms (protocols) the local voting protocol with nonvanishing step size was proposed. Nonvanishing (e.g., constant) step size ensures better transients in the time-invariant case and provides bounded error in the case of time-varying loads and agent states. The price to pay is replacement of the mean square convergence with an approximate one.

Analytic conditions for approximate consensus in stochastic network with noise, delays and switched topology were proposed. These conditions are based on the method of averaged models. This method allows to reduce the complexity of the closed loop system analysis. In this paper, upper bounds for the mean square distance between the initial system and its approximate average model were proposed. The proposed upper bounds were used to obtain conditions for approximate consensus achievement. In contrast to our previous works, we relaxed the assumption of the weights boundedness of the protocol replacing it by the boundedness of its variances.

%The theoretical results are applied to the balancing problem of information capabilities in sensor network. Theoretical results are confirmed by simulation.
The theoretical results were applied to the load balancing problem in a stochastic network. Theoretical results were confirmed analytically and by simulation. The large size simulation experiments were performed for the stochastic computer network. They showed that the performance of the adaptive multi-agent strategy with the redistribution of jobs among ``connected'' neighbors is significantly better than the performance of the strategy without redistribution.

% or


\appendices

%\section{Proof of Proposition \ref{Prop1}}

%\section{Proof of Proposition \ref{Prop2}}

%\section{Proof of Proposition \ref{Prop3}}

%\section{Proof of Proposition \ref{Prop3n}}

\section{Proof of Theorem \ref{Nat_L11}}
\begin{IEEEproof}
The following facts will be useful, for the remainder.
\begin{Proposition}\label{Prop1}
For $\bar z \in {\mathbb R}^n$ and matrix $A_{\max}$ the following inequality  holds
\begin{equation}
\sum_{i=1}^n (\sum_{j\in  N^{i}_{\max}} a_{\max}^{i, j} z^j)^2 \leq ||A_{\max}||^2_2||\bar z||^2.
\end{equation}
\end{Proposition}
\begin{proof}
Using the Cauchy-Schwarz inequality we obtain
\begin{equation}
\sum_{i=1}^n (\sum_{j\in  N^{i}_{\max}} a_{\max}^{i, j} z^j)^2 \leq \sum_{i=1}^n (\sum_{j\in  N^{i}_{\max}} {a_{\max}^{i, j}}^2)(\sum_{j\in  N^{i}_{\max}} {z^j}^2) \leq \end{equation}
$$ \leq (\sum_{i=1}^n \sum_{j=1}^n {a_{\max}^{i, j}}^2)(\sum_{j=1}^n {z^j}^2) \leq ||A_{\max}||^2_2||\bar z||^2.
$$
\end{proof}

For $t = 1, 2, \ldots $ we define an increasing sequence of $\sigma$-algebras $\tilde{\cal F}_{t}$ of probability events, generated by random elements $A_1, \ldots, A_{t-1};$ $ d_1^{i, j},$ $ \ldots, d_{t-1}^{i, j},$ $ b_1^{i, j},$ $ \ldots, b_{t-1}^{i, j}, w_1^{i, j},$ $ \ldots, w_{t}^{i, j},$ $i, j \in N$, and
${\cal F}_t=\sigma\{{\cal F}_t^A, A_{t}; b_{t}^{i, j}, d_{t}^{i, j}, i, j \in N\}$.

%For a random variable $Q$ and $\sigma$-algebra of probability event ${\cal F}$ we use the notation $\mathrm{E}_{\cal F} Q$ for the conditional expectation $Q$ with respect to $\sigma$-algebra ${\cal F}$.

Note that the random variables $\bar x_t$ are measurable with respect  $\sigma$-algebra ${\cal F}_{t-1}$, i.e. $\mathrm{E}_{{\cal F}_{t-1}} \bar x_t = \bar x_t$.

\begin{Proposition}\label{Prop2}
\begin{equation}||\bar s(\bar z)||^2 \leq 2 ||{\cal L}(A_{\max})||_2^2 ||\bar z||^2.
\end{equation}
\end{Proposition}
\begin{proof}
Using the result of Proposition~\ref{Prop1} yields
\begin{equation}||\bar s(\bar z)||^2=
\sum_{i=1}^n ( \sum_{j=1}^n a_{\max}^{i, j}( z^j-z^i))^2 \leq \sum_{i=1}^n ( d^i(A_{\max})|z^i|+
\end{equation}
$$+ |\sum_{j\in  N^{i}_{\max}} a_{\max}^{i,j} z^j|)^2 \leq 2( \sum_{i=1}^n d^i(A_{\max})^2 +  ||A_{\max}||_2^2) ||\bar z||^2 =
$$
$$
2 ||{\cal L}(A_{\max})||_2^2 ||\bar z||^2.
$$
\end{proof}

\begin{Proposition}\label{Prop3} If {\bf A2} is satisfied then $s^i(\bar x) = \frac{1}{\alpha_t}\mathrm{E}_{{\cal F}_{t-1}} {u_t^i}$ and the following inequality holds
\begin{equation}\frac{1}{\alpha_t^2}\mathrm{E}_{{\cal F}_{t-1}} {u_t^i}^{2}  \leq (||\bar x_t - x_t^i \underline 1 ||^2 + \sigma_w^2) \bar b, \; i \in N.
\end{equation}
\end{Proposition}

\begin{IEEEproof}
By the definition of the protocol~\eqref{Nat_7}
\begin{equation}
\frac{1}{\alpha_t} u_t^i = \mathop{\sum }\limits_{j\in \bar N_{t}^{i} } b_{t}^{i, j} ((x^j - x^i)+ (w_{t}^{i, j}  - w_{t}^{i, i} )).
\end{equation}
It follows from conditions {\bf A2} that $s^i(\bar x) = \frac{1}{\alpha_t}\mathrm{E}_{{\cal F}_{t-1}} {u_t^i}$.

By the centrality of observation noise (on condition {\bf {\bf A2a}}) we consecutively derive
\begin{equation}
\frac{1}{\alpha_t^2}\mathrm{E}_{{\cal F}_{t-1}} {u_t^i}^{2} = \mathrm{E}_{{\cal F}_{t-1}}  (\mathop{\sum }\limits_{j\in \bar N_{t}^{i} } b_{t}^{i,j} ((x_t^j - x_t^i)+ (w_{t}^{i,j}  - w_{t}^{i,i} )))^2 =
\end{equation}
$$
= \mathrm{E}_{{\cal F}_{t-1}}  (\mathop{\sum }\limits_{j\in \bar N_{t}^{i} } b_{t}^{i,j} (x_t^j - x_t^i))^2+
 \mathrm{E}_{{\cal F}_{t-1}}  (\mathop{\sum }\limits_{j\in \bar N_{t}^{i} }b_{t}^{i,j} (w_{t}^{i,j}  - w_{t}^{i,i} ))^2 \leq
 $$
$$ \leq ||\bar x_t - x_t^i \underline 1 ||^2 \mathrm{E}_{{\cal F}_{t-1}}  \mathop{\sum }\limits_{j\in \bar N_{t}^{i} } (b_{t}^{i,j})^2
 + \mathrm{E}_{{\cal F}_{t-1}}  \mathop{\sum }\limits_{j\in \bar N_{t}^{i} } {b_{t}^{i,j}}^2 ({w_{t}^{i,j}}^2 +
 $$
 $$+{ w_{t}^{i,i}}^2) \leq (||\bar x_t - x_t^i \underline 1 ||^2 + \sigma_w^2) \bar b.
$$
\end{IEEEproof}

\begin{Proposition}\label{Prop3n} If {\bf A2} is satisfied then
the following inequality holds
\begin{equation}\mathrm{E}_{{\cal F}_{t-1}}|\frac{1}{\alpha_t}u_t^i - s^i(\bar x)|^{2}  \leq (n-1) \bar b^2  ||\bar x_t - x_t^i \underline 1 ||^2 + n \bar b^2 \sigma_w^2, \; i \in N.
\end{equation}
\end{Proposition}

\begin{IEEEproof}
The conclusion of Proposition~\ref{Prop3n} follows from Proposition~\ref{Prop3} since
\begin{equation} \mathrm{E}_{{\cal F}_{t-1}}|\frac{1}{\alpha_t}u_t^i - s^i(\bar x)|^{2} = (\mathrm{E}_{{\cal F}_{t-1}}\frac{1}{\alpha_t}u_t^i)^2 - s^i(\bar x)^2 \leq (\mathrm{E}_{{\cal F}_{t-1}}\frac{1}{\alpha_t}u_t^i)^2.
\end{equation}

\end{IEEEproof}

To proof the Theorem~\ref{Nat_L11} we need to show that the Lipschitz and growth conditions from \cite{fradkov} hold.
The first is a direct consequence of the Lipschitz continuous function $f^i(x, u)$ and the form of vector function $R(\alpha, \bar z)$.
Let $\bar z, \bar z' \in {\mathbb R}^n$.
By Proposition~\ref{Prop1} we have
\begin{equation}||R(\alpha, \bar z) - R(\alpha, \bar z')|| =
\left(
\frac{L_1^2}{\alpha^2}\sum_{i=1}^n (L_x|z^i-{ z'}^i| +  |\alpha   s^i(\bar z -\bar z')|)^2
\right)^{\frac{1}{2}}
\leq
\end{equation}
$$\leq L_1 \sqrt{2} \sqrt {\frac{L_x}{\alpha^2}+ 2||{\cal L}(A_{\max})||_2^2 } ||(\bar z-\bar z')|| = \bar L_1 ||\bar z-\bar z'||.
$$
Similarly
\begin{equation}
||R(\alpha, \bar z) - R(\alpha', \bar z)|| =
\end{equation}
$$
\left(\sum_{i \in N} (\frac{1}{\alpha} f^i(z^i, \alpha s^i(\bar z)) - \frac{1}{\alpha'} f^i(z^i, \alpha' s^i(\bar z)))^2\right)^{\frac{1}{2}} =
$$
$$ = (\sum_{i \in N} (\frac{1}{\alpha} (f^i(z^i, \alpha s^i(\bar z)) - f^i(z^i, \alpha' s^i(\bar z))) -
$$
$$ - (\frac{1}{\alpha'} - \frac{1}{\alpha}) f^i(z^i, \alpha' s^i(\bar z)))^2)^{\frac{1}{2}} \leq (2\sum_{i \in N} \frac{L_1^2(\alpha - \alpha')^2}{\alpha^2} | s^i(\bar z)|^2 +
$$
$$
L_2(\frac{1}{\alpha'} - \frac{1}{\alpha})^2(L_c+L_x|z^i|^2 + {\alpha'}^2| s^i(\bar z)|^2))^{\frac{1}{2}} \leq \bar L_{\alpha}(1 + ||\bar z||)|\alpha - \alpha'|.
$$

Next let us prove that the growth condition. Let $\bar z \in {\mathbb R}^n$.
Due to the limited growth rate $f^i(x,u)$ and Lipschitz property in $u$ (by assumption {\bf A1})
we have
\begin{equation}{ \mathrm{E}}||\frac{1}{\alpha_t}F(\alpha_t, \bar z, \bar w_{t}) - R(\alpha, \bar z)||^2 = \sum_{i \in N} \mathrm{E}|\frac{1}{\alpha_t} f^i(z^i, \alpha_{t} \tilde s_t^i) -
\end{equation}
$$ - \frac{1}{\alpha} f^i(z^i, \alpha s^i(\bar z))|^2 = \sum_{i \in N} \mathrm{E}|\frac{1}{\alpha_t}( f^i(z^i, \alpha_{t} \tilde s_t^i) - f^i(z^i, \alpha s^i(\bar z))) - $$
$$- (\frac{1}{\alpha} - \frac{1}{\alpha_t}) f^i(z^i, \alpha s^i(\bar z))|^2 \leq 2 \sum_{i \in N} \mathrm{E}  \frac{1}{\alpha_t^2} |f^i(z^i, \alpha_{t} \tilde s_t^i) -
$$
$$
- f^i(z^i, \alpha s^i(\bar z))|^2 +  (\frac{1}{\alpha} - \frac{1}{\alpha_t})^2 |f^i(z^i, \alpha s^i(\bar z))|^2 \leq
$$
$$\leq 2 \sum_{i \in N} \mathrm{E} \frac{L_1}{\alpha_t^2} |\alpha_t \tilde s_t^i - \alpha s^i(\bar z)|^2 +  L_2(\frac{1}{\alpha} - \frac{1}{\alpha_t})^2  (L_c+L_x|z^i|^2 +
$$
$$+|s^i(\bar z)|^2)\leq \gamma_t (n L_c+L_x||\bar z||^2)+\sum_{i \in N} \mathrm{E} 4 L_1 |\tilde s_t^i - s^i(\bar z)|^2 +
$$
$$
+ (\frac{4L_1({\alpha} - {\alpha_t})^2}{\alpha_t^2} + \gamma_t ) s^i(\bar z)^2,
$$
where
$\gamma_t = 2 L_2(1/{\alpha} - 1/{\alpha_t})^2 $.

Denote $\beta_t = (4 L_1 (({\alpha} - {\alpha_t})^2-1)/\alpha_t^2+\gamma_t$ and by Propositions~\ref{Prop2}, \ref{Prop3}, \ref{Prop3n} consistently derive
\begin{equation}{ \mathrm{E}}||\frac{1}{\alpha_t}F(\alpha_t,  \bar z, \bar w_{t}) - R(\alpha, \bar z)||^2 \leq \gamma_t (nL_c+L_x||\bar z||^2) +\end{equation}
$$
+\sum_{i \in N} (4 \frac{L_1}{\alpha_t^2} (\mathrm{E}(\tilde s_t^i)^{2} - s^i(\bar z)^2) + (\frac{4L_1({\alpha} - {\alpha_t})^2}{\alpha_t^2} + \gamma_t ) s^i(\bar z)^2 =
$$
$$ = \gamma_t (nL_c+L_x||\bar z||^2)+4 \frac{L_1}{\alpha_t^2} \mathrm{E}||{\tilde s_t}||^{2} + \beta_t  ||\bar s(\bar z)||^2 \leq
$$
$$ \leq n \gamma_t L_c + 4 \frac{L_1}{\alpha_t^2} n^2 \bar b^2 \sigma_w^2 +
(\gamma_t L_x + 8n(n-1) \bar b^2  \frac{L_1}{\alpha_t^2} +
$$
$$+ 2 \beta_t  ||{\cal L}(A_{\max})||_2^2) ||\bar z||^2 =\bar L_2 (1+||\bar z||^2).
$$
\end{IEEEproof}

\section{Proof of Theorem \ref{Nat_T11}}
\begin{IEEEproof}
To proof Theorem~\ref{Nat_T1}, the following facts will be useful.
%При доказательстве Теоремы~\ref{Nat_T1} будут использованы следующие четыре вспомогательных предложения:

%{\it Лемма П.4:}
\begin{Proposition}\label{Prop4}
\begin{equation}
||U\bar X||^2 \leq 2^{\tilde d }||\bar X||^2, \; \ldots,
\;||U^{\bar d} \bar X||^2 \leq 2^{\bar d} ||\bar X||^2, \; \ldots,
\;||U^k \bar X||^2 \leq
\end{equation}
$$\leq 2^{\bar d} ||\bar X||^2,$$\end{Proposition}

%{\it Д~о~к~а~з~а~т~е~л~ь~с~т~в~о:}
\begin{IEEEproof}
By the definition of matrix $U$ it is easy to obtain the first inequality, and the rest we get by induction on $k$ and by the following equality
\begin{equation}
\label{Nat_9_barD}
\forall k>\bar d\;\;U^k = U^{\bar d} =  \left(\begin{array}{ccccc} {I} & {0} & {0} & {\ldots } & {0} \\ {I} & {0} & {0} & {\ldots } & {0}
%\\ {I} & {0} & {0} & {\ldots } & {0}
\\ {\vdots } & {\vdots } & {\vdots } & {\vdots } & {\vdots } \\
{I} & {0} & {0} & {\ldots } & {0} \end{array}\right).
\end{equation}
\end{IEEEproof}

Denote
\begin{equation}
\label{Nat_10_ht}
v_t = F(\alpha_t, \bar  X_{t}, \bar w_{t}) - G(\alpha_{t}, \bar X_{t}).
 \end{equation}

%{\it Лемма П.1:}
\begin{Proposition}\label{Prop1F}
%При выполнении условий {\bf A2} справедлива оценка
By assumptions {\bf A2} the following inequality holds
\begin{equation}
\mathrm{E} \max_{1\leq t \leq T} ||\sum_{i=1}^t v_{t} ||^2 \leq 4 n  \sum_{t=1}^T \mathrm{E} ||v_{t} ||^2.
\end{equation}
\end{Proposition}
\begin{IEEEproof}
Under the conditions {\bf A2} random elements $v_t$ are martingale differences, i.e., they are centered with respect to the conditional averaging of the background: $\mathrm{E}_{{\cal F}_{t-1}} v_t = 0$. So, Lemma~1 from section~3 of~\cite{Gihman} is applicable. The dimension of vectors $v_t$ is $n \bar d$, but since only the first $n$ components of vectors $v_t$ are nonzero, then it is possible to use in the estimation the value of~$n$ instead of $n \bar d$.
\end{IEEEproof}

%При выполнении условий {\bf A2}  случайные элементы $v_t$ являются мартингальными разностями, т.~е. они центрированы при условном усреднении относительно предыстории: $\mathrm{E}_{{\cal F}_{t-1}} v_t = 0$. Следовательно, применима Лемма~1 из параграфа~3 работы~\cite{Gihman}. Размерность векторов $v_t$ равна $n \bar d$, но так как только первые $n$ компонент векторов $v_t$ отличны от нуля, то в формуле для оценки можно вместо $n \bar d$ можно использовать величину~$n$.

%{\it Лемма П.3}:
\begin{Proposition}\label{Prop3F}
%Пусть последовательность чисел $\mu_t \geq 0$, $t=0, 1, \ldots, T$ удовлетворяет неравенствам
Let the sequence of numbers $\mu_t \geq 0$, $t=0, 1, \ldots, T$ satisfies the inequalities
\begin{equation}
\mu_{t+1} \leq \bar  \alpha c_1 \tau_t + c_2 2^{\bar d} \tau_t \sum_{k=1}^t \gamma_k  \mu_k,\; c_1, c_2 \ge 0,\;
\end{equation}
then
\begin{equation}
\mu_t \leq  c_1 \tau_t  e^{ c_2 \tau_t^2 } \bar  \alpha.
\end{equation}
\end{Proposition}
\begin{IEEEproof}
Statement of Proposition follows directly from the corresponding result in \cite{Bernshtein64}
 \end{IEEEproof}

% Утверждение предложения непосредственно следует из соответствующего результата в \cite{Bernshtein64}.% при подстановке

%{\it Лемма П.2:}
\begin{Proposition}\label{Prop2F}
By assumptions {\bf A1}, {\bf A2} yields
\begin{equation}
\mathrm{E} ||\bar X_{t}||^2 \leq (\frac{2 n  L_2  + \bar \alpha^2 \tilde c}{c_3}+||\bar X_0||^2) e^{t\ln (c_3+1)}.
\end{equation}
\end{Proposition}
\begin{IEEEproof}
We write equation~\eqref{Nat_8} as
 \begin{equation}
\label{Nat_8_D23}
\bar X_{t+1} = U  \bar X_{t} + G(\alpha_{t}, \bar X_{t}) + v_t.
\end{equation}
For the squared norm of $\bar X_{t+1}$ we have
 \begin{equation}
\label{Nat_82_D}
||\bar X_{t+1}||^2 = ||U  \bar X_{t} + G(\alpha_{t}, \bar X_{t})||^2 + 2 (U  \bar X_{t} + G(\alpha_{t}, \bar X_{t}))^{\rm T} v_t + ||v_t||^2.
\end{equation}

Taking the conditional expectation of both parts of (\ref{Nat_82_D}) on $\sigma$-algebra ${\cal F}_{t-1}$ (i.e. for fixed $\bar X_t$) by the centrality of $v_t$ we obtain
$$
\mathrm{E}_{{\cal F}_{t-1}}{||\bar X_{t+1}||^2} = ||U  \bar X_{t} + G(\alpha_{t}, \bar X_{t})||^2 +  \mathrm{E}_{{\cal F}_{t-1}}||v_t||^2 \leq
$$
 \begin{equation}
\label{Nat_8_D_2}
\leq 2||U  \bar X_{t}||^2 + 2|| G(\alpha_{t}, \bar X_{t})||^2 +  \mathrm{E}_{{\cal F}_{t-1}}||v_t||^2.
\end{equation}

By the form of $v_t$ and Lipschitz in $u$ of functions $f^i(u)$ (by {\bf A1}) for  $||v_{t}||^2$  we have
 \begin{equation}
||v_{t}||^2 = \sum_{i \in N} |f^i(x_t^i, \alpha_{t} \mathop{\sum }\limits_{j\in \bar N_{t}^{i} } b_{t}^{i, j} (x_{t-d_t^{i, j}}^j - x_t^i+ w_{t}^{i, j}  - w_{t}^{i, i} ))  -
 \end{equation}
$$
- f^i(x_t^i, {\alpha_t } s^i_t(\bar X_t))|^2 \leq
  L_1^2  ||\bar u_t-\alpha_{t}^2\bar s_t||^2.
 $$

Under the conditions {\bf A2}, random variables
$\mathrm{E}_{{\cal F}_{t-1}} u^i_t$, $i \in N$ satisfy the conditions of Proposition~\ref{Prop3}
 \begin{equation}
\label{Nat_8_D26}
\mathrm{E}_{{\cal F}_{t-1}} ||v_{t}||^2 = \alpha_t^2 L_1^2(2n \bar b||\bar X_t||^2+n^2  \bar b \sigma_w^2).
\end{equation}

Consistently evaluating all three summands on the right hand side of~(\ref{Nat_8_D_2}) and taking  into account the results of Propositions~\ref{Prop4}, \ref{Prop2} and \ref{Prop3}, we deduce
 \begin{equation}
\mathrm{E}_{{\cal F}_{t}}{||\bar X_{t+1}||^2} \leq
 2^{\tilde d}  ||\bar X_{t}||^2 + 2^{1+\tilde d/2}  ||\bar X_{t}||  L_1 (L_x||\bar X_{t}||+{\alpha_t}||\bar s ||)      +
 \end{equation}
 $$
+ L_2 (n L_c+L_x||\bar X_{t}||^2+{\alpha_t^2}||\bar s ||^2)
   +
 \alpha_t^2 L_1^2(2n \bar b||\bar X_t||^2+ $$
  $$
 +2 n \bar b \sigma_w^2) \leq (2^{\tilde d}+2^{1+\tilde d/2}L_1 L_x+L_2 L_x+\alpha_t 2^{1+\tilde d/2}L_1||{\cal L}(A_{\max})||_2 +
 $$
 $$+{\alpha_t^2}(L_2||{\cal L}(A_{\max})||_2^2 +
 2n L_1^2 \bar b))||\bar X_t||^2
+ n L_2 L_c +
$$
$$
+ 2{\alpha_t^2} n L_1^2 \bar b \sigma_w^2 \leq \bar c + \bar c_3 ||\bar X_{t}||^2,
$$
%\end{equation}
where $\bar c = n L_2 L_c + {\alpha_t^2} \tilde c,\;\bar c_3 = c_3 +1.$

By taking unconditional expectation of both parts of this inequality and consistently iterating on $t$, we obtain Proposition~\ref{Prop2F}
 \begin{equation}
\mathrm{E}||\bar X_{t}||^2 \leq \bar c + \bar c_3 \mathrm{E}||\bar X_{t-1}||^2 \leq \bar c +  \bar c \bar c_3 + \bar c_3^2 \mathrm{E}||\bar X_{t-2}||^2 \leq
 \end{equation}
$$
\leq  \bar c (1 + \bar c_3 + \bar c_3^2 + \ldots + \bar c_3^{t-1}) + \bar c_3^t ||\bar X_0||^2 \leq  \bar c \frac{\bar c_3^t-1}{c_3} + \bar c_3^t ||\bar X_0||^2 \leq
$$
$$
\leq \left(\frac{ \bar c}{c_3} + ||\bar X_0||^2\right)\bar c_3^t \leq (\bar c_4 + ||\bar X_0||^2) e^{t\ln \bar c_3},
$$
where $\bar c_4 = {\bar c}/{c_3}$.
\end{IEEEproof}

Denote $x^{\star}$ as the consensus value of the continuous model.
From the first group of conditions of Theorem~\ref{Nat_T11} the conditions of Theorem~\ref{Nat_L11} hold, i.e. the result of the theorem is true. From other conditions of Theorem~\ref{Nat_T11} and the result of Theorem~\ref{Nat_L11} we obtain
\begin{equation}
\mathrm{E}||\bar x_t - x^{\star}\underline 1||^2 \leq 2 \mathrm{E}||\bar x_t - \bar x(\tau)||^2 + 2 ||\bar x(\tau) - x^{\star}\underline 1||^2 \leq \frac{\varepsilon}{2}+\frac{\varepsilon}{2}\leq \varepsilon.
\end{equation}
\end{IEEEproof}

\section{Proof of Corollary \ref{Nat_T22}} %JYM: Collorary
\begin{IEEEproof}
In the conditions of Corollary~\ref{Nat_T22}
$R(\alpha, \bar x)$ is a linear function. Therefore the dynamical system \eqref{e2} takes the form:
\begin{equation}
\dot {\bar x} = -  {\cal L}(A_{\max}) \bar x,
\end{equation}
where ${\cal L}(A_{\max})$ is the Laplacian of $A_{\max}$.
All amounts in rows of elements of the matrix ${\cal L}(A_{\max})$ are equal to zero and, moreover, all the diagonal elements are positive and equal to the absolute value of the sum of all the other elements in the row.
The vector of 1's $\underline 1$ is the right eigenvector corresponding to zero eigenvalue.
The resulting continuous system is partially stable with respect to $h=\underline 1^T \bar x$.

By condition {\bf A3} it was obtained that in this continuous system the asymptotic consensus is achieved, and in Lemma~\ref{Nat_ContConsTime} the $\varepsilon$-consensus is achieved, and the time to consensus is given by \eqref{NL_Teps}.

\end{IEEEproof}

%\section{Proof of Proposition\ref{Prop4}}

%\section{Proof of Proposition\ref{Prop1F}}

%\section{Proof of Proposition\ref{Prop3F}}

%\section{Proof of Proposition\ref{Prop2F}}

\section{Proof of Theorem \ref{Nat_T1}}
\begin{IEEEproof}
  By condition {\bf A2} averaging with respect to $\sigma$-algebras ${\cal F}_{t}^d$ and ${\cal F}_{t}$ yields $\mathrm{E}_{{\cal F}_t} v_t = 0$.
By iterating equation~\eqref{Nat_8} for $t, t-1, \ldots t-d+1$ we obtain
$$
\bar X_{t+1} = U  \bar X_{t} + G(\alpha_{t}, \bar X_{t}) +  v_t =
$$
 \begin{equation}
\label{Nat_8_D}
= U^2  \bar X_{t-1} + U G(\alpha_{t-1},\bar X_{t-1}) + G(\alpha_{t}, \bar X_{t}) +  U v_{t-1}+  v_t =
\end{equation}
$$
= \cdots = U^{t+1}  \bar X_{0} + \sum_{k=0}^t U^{t-k} G(\alpha_{k}, \bar X_{k})+\sum_{k=0}^t  U^{t-k}  v_k.
$$
Similarly we obtain
 \begin{equation}
\label{Nat_8_D_23}
\bar Z_{t+1} =  U^{t+1}  \bar X_{0} + \sum_{k=0}^t  U^{t-k} G(\alpha_{k}, \bar Z_{k}).
\end{equation}
Let us estimate $||\bar X_{t}-\bar Z_{t}||^2,\; t = 1, \ldots,T$. By subtracting (\ref{Nat_8_D_23}) from (\ref{Nat_8_D}) and squaring the result we obtain $$||\bar X_{t}-\bar Z_{t}||^2 =|| \sum_{k=1}^t  U^{t-k} v_k+\sum_{k=1}^t  U^{t-k} (G(\alpha_{k}, \bar X_{k})-G(\alpha_{k}, \bar Z_{k}))||^2 \leq $$
$$
\leq   2  || \sum_{k=1}^t  U^{t-k} v_k||^2 + 2 ||\sum_{k=1}^t  U^{t-k} (G(\alpha_{k}, \bar X_{k})-G(\alpha_{k}, \bar Z_{k}))||^2 \leq
$$
 \begin{equation}
\label{Nat_8_D31}
%\leq
\leq 2  || \sum_{k=1}^t  U^{t-k} v_k||^2 + 2 \frac{\tau_t}{2^{\bar d}} \sum_{k=1}^t \frac{1}{\alpha_t} || U^{t-k} (G(\alpha_{k}, \bar X_{k})-G(\alpha_{k}, \bar Z_{k}))||^2.
\end{equation}

For the summands in the second sum of~\eqref{Nat_8_D31} using Propositions~\ref{Prop2}, \ref{Prop4} and Lipschitz condition $f^i(\cdot, \cdot)$ (assumption {\bf A1}) we obtain
 \begin{equation} || U^{t-k} (G(\alpha_{k}, \bar X_{k})-G(\alpha_{k}, \bar Z_{k}))||^2 \leq 2^{\bar d} L_1^2 \sum_{i=1}^n(L_x |x^i_k - z^i_k| +  \end{equation}
$$
+\alpha_k|s(x^i_k)-s(z^i_k)|)^2 \leq 2^{1+\bar d} L_1^2 \sum_{i=1}^n L_x |x^i_k - z^i_k|^2 +
\alpha_k^2 s(x^i_k-z^i_k)^2\leq
$$
$$
\leq
2^{1+\bar d} L_1^2 (L_x+2\alpha_k^2 ||{\cal L}(A_{\max})||_2^2) || \bar X_{k} -\bar Z_{k} ||^2
$$

We take expectation of both parts of \eqref{Nat_8_D31} and denote $\mu_T = \mathop{\max }\limits_{0\le {t} \le T } \mathrm{E} || \bar X_{t} -\bar Z_{t} ||^2$. By applying Proposition~\ref{Prop1F} to the first summand and obtained above estimate of the second summand we obtain
 \begin{equation}
\label{Nat_8_D32}
\mu_T \leq 2^{3+\bar d} n   \sum_{k=1}^T  \mathrm{E} ||v_k||^2 +2\tau_T  L_1^2 \sum_{k=1}^t (\frac{L_x}{\underline \alpha}+2 \alpha_k||{\cal L}(A_{\max})||_2^2)    \mu_k.
\end{equation}
To estimate $\mathrm{E} ||v_k||^2$ by using previously obtained relation~(\ref{Nat_8_D26}) and the result of Proposition~\ref{Prop2F} we deduce
 \begin{equation} \mathrm{E}||v_{k}||^2 \leq \alpha_{k}^2  (\tilde c  + \hat c (\bar c_4+||\bar X_0||^2) e^{k\ln (c_3+1)})
 \end{equation}
and hence
 \begin{equation}
\label{Nat_8_D33}
2^{3+\bar d} n   \sum_{k=1}^T  \mathrm{E} ||v_k||^2 \leq \bar \alpha 8  n \tau_T (\tilde c  + \hat c (\bar c_4+||\bar X_0||^2) e^{T\ln (c_3+1)}).
\end{equation}

By the following relation
$
2^{\bar d}\sum_{k=1}^t \alpha_{k}^2 \leq \bar \alpha 2^{\bar d}\sum_{k=1}^t \alpha_{k} = \bar \alpha \tau_t,
$
considering estimates~(\ref{Nat_8_D33}) from (\ref{Nat_8_D32}), we have
 \begin{equation}
\label{Nat_8_D34}
\mathrm{E} \mu_T \leq \bar  \alpha  c_1 \tau_T + c_2\tau_T 2^{\bar d} \sum_{k=1}^T \alpha_{k} \mathrm{E} \mu_k.
\end{equation}

From last inequality~(\ref{Nat_8_D34}) by applying Proposition~\ref{Prop3F} we get the conclusion of Theorem~\ref{Nat_T1}.
\end{IEEEproof}

\section{Proof of Theorem \ref{Nat_T12}}
\begin{IEEEproof}
 Denote $x^{\star}$ as consensus value of discrete system~\eqref{Nat_10_Zt}. From the first group of conditions of Theorem~\ref{Nat_T12} the conditions of Theorem~\ref{Nat_T1} hold. From other conditions of Theorem~\ref{Nat_T12} and the result of Theorem~\ref{Nat_T1} we obtain
 \begin{equation}
\mathrm{E}||\bar X_t - x^{\star}\underline 1||^2 \leq 2 \mathrm{E}||\bar X_t - \bar Z_t||^2 + 2 ||\bar Z_t - x^{\star}\underline 1||^2 \leq \frac{\varepsilon}{2}+\frac{\varepsilon}{2}\leq \varepsilon.
 \end{equation}
\end{IEEEproof}

\section{Proof of Theorem \ref{Nat_T22new}}
\begin{IEEEproof}
The result of Theorem~\ref{Nat_T22new} is derived from Theorem~\ref{Nat_T12}.

All amounts in rows of elements of the matrix $\bar {\cal L} = (I-U) -{\cal L}( \alpha A_{\max}) $ are equal to zero and, moreover, all the diagonal elements are positive and equal to the absolute value of the sum of all the other elements in the row, which are negative. Hence the matrix $\bar {\cal L}$ is the Laplacian of a graph and a vector of 1's $\underline 1$ is the right eigenvector corresponding to zero eigenvalue.

By condition {\bf A3}, the graph corresponding to the Laplacian $\bar {\cal L}$ has a spanning tree. By condition~{\bf A3} graph of the first $n$ nodes has a spanning tree. And units on $(n+1)$-th diagonal consistently connect $\bar n$-th node with $(\bar n-\bar d)$-th node, $(\bar n-1)$-th node with $(\bar n-\bar d-1)$-th and so on. Hence the asymptotic consensus is achieved in such a discrete system since the condition $\alpha < \frac{1}{d_{\max}}$ holds by the assumptions of Theorem~\ref{Nat_T22new}.

To satisfy the conditions of Theorem~\ref{Nat_T12} it remains to show that the constants $\bar C_1$ and $\bar C_2$ are the same as the corresponding constants from Theorem~\ref{Nat_T1}. It follows from the fact that in this case $L_1=L_2=1$,  $L_x=L_c=0$.
\end{IEEEproof}

\section{Proof of Lemma \ref{lemma_optim}}
% \textit{\textbf{(Of the optimal control strategy.)}}
\begin{proof}

We take $x_t^i=q_t^i/p_t^i$ as the state of agent $i$.

We give the proof by contradiction. Assume that for some optimal strategy not all $x_t^i$ are equal to each other, i.e. there is a agent with the number $k \in N$ and the subset of agents  $\tilde N_t$ such that $x_t^k  > x_t^j, \; \forall j \;\in \tilde N_t$.

Denoted by $l=|\tilde N_t|$ the number of agents in $\tilde N_t$. The states of other $n-l$ agents equal $x_t^k$.

Let the difference between the state of $k$-th agent and the biggest of the set $\tilde N_t$ equals to $\epsilon_t$, i.e.
\begin{equation}
\epsilon_t = x_t^k - \max_{ j \;\in \tilde N_t } x_t^j.
\end{equation}

%Рассмотрим новую стратегию распределения заданий.
Let's consider the new strategy of job redistribution. Reduce the load of all $n-l$ agents which have the maximum load on $\frac{\epsilon}{2(n-l)}$ (i.e. on $\frac{\epsilon}{2}$ at all) and add this $\frac{\epsilon}{2}$ jobs to any of the $l$ agents of $\tilde N_t$. For new strategy we found that the time of job redistribution in the system will be less than the initial on the $\frac{\epsilon}{2(n-l)}$, i.e. less than the minimum by the assumption. A contradiction.

%Уменьшим загрузку всех $n-l$ узлов, имевших максимальную загрузку, на $\frac{\epsilon}{2(n-l)}$  (т.~е. всего на $\frac{\epsilon}{2}$) и добавим эти $\frac{\epsilon}{2}$ заданий к любому из $l$ узлов из $\tilde N_t$. Для новой стратегии получили, что время обработки заданий в системе будет меньше исходного на $\frac{\epsilon}{2(n-l)}$, т.~е. меньше, чем минимальное по предположению. Получили противоречие.

\end{proof}

\section{Proof of Theorem \ref{Nat_T3}}
\begin{IEEEproof}
You should verify if the conditions {\bf A1}, {\bf A2} for the considered control protocol and functions $f^i(\cdot,\cdot)$ are satisfied. If they are satisfied, then all the conditions of Theorem~\ref{Nat_T12} are satisfied and the result is valid for this case.

The condition {\bf A1} holds since the function $f^i(x,u) = -1 +u$  is linear in $u$.
The condition {\bf A2} holds because of the formation rules for the weighting coefficients in the control protocol and stabilization conditions for~$p_t^i$.

\end{IEEEproof}
%\section{}

\section*{Acknowledgment}


This work was carried out during the tenure of an ERCIM ``Alain
Bensoussan'' Fellowship Programme. The research leading to these
results has received funding from the European Union Seventh Framework
Programme (FP7/2007-2013) under grant agreement No. 246016.
Also the work was supported by Russian Federal Program ``Cadres'' (agreements 8846, 8855), by RFBR (project 11-08-01218, 13-07-00250), and by the SPRINT laboratory of SPbSU and Intel Corp.

\ifCLASSOPTIONcaptionsoff
  \newpage
\fi



%\bibliographystyle{IEEEtran}

\bibliography{NatRef}
\bibliographystyle{IEEEtran}

%\begin{biography}[{\includegraphics[width=1in,height=1.25in,clip,keepaspectratio]{mshell}}]{Michael Shell}

%\begin{IEEEbiography}{Natalia~Amelina}
%\end{IEEEbiography}

%\begin{IEEEbiography}{Alexander~Fradkov}
%\end{IEEEbiography}

%\begin{IEEEbiography}{Yuming~Jiang}
%\end{IEEEbiography}

%\begin{IEEEbiography}{Dimitrios~J.~Vergados}
%\end{IEEEbiography}


%\newpage



%\vfill


\end{document}